\documentclass[aps,pre,twocolumn,showpacs]{revtex4}
\usepackage{graphicx}
\usepackage{amsmath,amscd,amssymb}
\usepackage{color}

\newcommand{\be}{\begin{equation}}
\newcommand{\ee}{\end{equation}}
\newcommand{\bea}{\begin{eqnarray}}
\newcommand{\eea}{\end{eqnarray}}

\newcommand{\br}{\mathbf{r}}

\newcommand{\e}{\varepsilon}

\newcommand{\tv}{\tilde{v}}

\newcommand{\brr}{\bar{r}}
\newcommand{\bbr}{\bar{\br}}

\newcommand{\bd}{\bar{d}}
\newcommand{\bk}{\bar{k}}

\begin{document}

\title{Electrostatic correlations on the ionic selectivity of cylindrical membrane nanopores}

\author{Sahin Buyukdagli$^{1,2}$\footnote{email:~\texttt{Sahin.Buyukdagli@iri.univ-lille1.fr}} and T. Ala-Nissila$^{2,3}$\footnote{email:~\texttt{Tapio.Ala-Nissila@aalto.fi}}}
\affiliation{$^{1}$Institut de Recherche Interdisciplinaire USR3078 CNRS and Universit\'e Lille I, Parc de la Haute Borne, 52 Avenue de Halley, 59658 Villeneuve d'Ascq, France\\
$^{2}$Department of Applied Physics and COMP Center of Excellence, Aalto University School of Science, P.O. Box 11000, FI-00076 Aalto, Espoo, Finland\\
$^{3}$Department of Physics, Brown University, Providence, Box 1843, RI 02912-1843, U.S.A.}
\date{\today}

\begin{abstract}
We characterize the role of electrostatic fluctuations on the charge selectivity of cylindrical nanopores confining electrolyte mixtures. To this end, we develop an extended one-loop theory that can account for correlation effects induced by the surface charge, nanoconfinement of the electrolyte, and interfacial polarization charges associated with the low permittivity membrane. We validate the quantitative accuracy of the theory by comparisons with previously obtained Monte-Carlo simulation data from the literature, and scrutinize in detail the underlying forces driving the ionic selectivity of the nanopore. In the biologically relevant case of electrolytes with divalent cations such as $\mathrm{CaCl}_2$ in negatively charged nanopores, electrostatic correlations associated with the dense counterion layer in the channel result in an increase of the pore coion density with the surface charge. This peculiarity analogous to the charge inversion phenomenon remains intact for dielectrically inhomogeneous pores, which indicates that the effect should be observable in nanofiltration membranes or DNA-blocked nanopores characterized by a low membrane permittivity. Our results show that a quantitatively accurate consideration of correlation effects is necessary to determine the ionic selectivity of nanopores in the presence of electrolytes with multivalent counterions.

\end{abstract}

\maketitle
\section{Introduction}
Electrostatic forces induced by the nanoscale confinement of charged liquids are at the origin of various industrial and biotechnological applications. A prominent example is the water desalination process that aims at separating salt from sea water by making use of the Donnan and dielectric exclusion mechanisms~\cite{yar1,yar2,Lang}. The coupling of electrostatic and hydrodynamic forces is also the basis of nanofluidics separation technics that consist in controlling the flow of charged liquids thorough nanochannels. Although a newborn science, nanofluidics has already found many important applications from electromechanical energy conversion~\cite{nano1} to DNA sequencing~\cite{nano2} and  polymerase chain reaction in lab-on-a-chip devices~\cite{nano3}. In order to predict and control the functioning of these devices, it is an important task to develop electrostatic formulations of confined charged liquids that can handle ion-ion and ion-substrate correlations in an accurate way; a challenge that remains to be met at present.

\textcolor{black}{One of the major limitations of the current nanofluidic transport theories is clearly the underlying dielectric continuum electrostatics. Field theoretic formulations incorporating the charge structure of solvent~\cite{dun1,orland1,buyukdagli13,jcpPOL} or fluctuating polyelectrolytes~\cite{dun2} have been previously proposed. The additional weak point of nanofluidic approaches is their mean-field (MF) nature.} More precisely, these formulations that couple hydrodynamic transport equations with the Poisson-Boltzmann (PB) equation neglect electrostatic correlations, which are known to be highly relevant for charged nanopores confining electrolytes of general composition~\cite{mcasy,mcmix}. On the side of the artificial nanofiltration studies, ion rejection models~\cite{yar1,yar2} that make use of correlation corrected formulations such as the electrostatic self-consistent (SC) equations~\cite{SCrussian,netzvar} have been so far one step ahead. However, uncontrolled approximations involved in the derivation of the SC equations and in their solution in cylindrical pore geometries have obscured the quantitative precision of these models.

The standard way to properly account for ionic correlation effects consists of the one-loop expansion of the electrolytic free energy around the PB theory. Unfortunately, the unavailability of analytical solutions for the one-loop level electrostatic equations in curved geometries have restricted the proper consideration of ionic correlation effects to simple planar systems, \textcolor{black}{whereas interfacial curvature effects have been mostly considered at the MF level~\cite{anan1,anan2,anan3}}. At one-loop order, electrostatic correlation effects on the interaction energy of charged plates were considered within a semiclassical approximation by Podgornik and Zeks~\cite{PodWKB} and in terms of special functions by Attard et al.~\cite{attard}. Correlation induced modifications of ion densities at charged planes were also investigated at the same perturbative order in Refs.~\cite{netzcoun,1loop,jcp2} or using strong-coupling expansion technics in Refs.~\cite{st1,st2,st3,st4}. However, ionic correlations in cylindrical geometries have been mostly studied within the DH approximation~\cite{cyl1,cyl2,cyl3}, which is more limited than the one-loop theory, and valid exclusively for neutral or very weakly charged cylinders where the surface charge induced electrostatic potential is much lower than the thermal energy $k_BT$.

The main advantage of the SC approach over the one-loop theory is its ability to deal with dielectrically inhomogeneous interfaces where the one-loop expansion is known to fail~\cite{David,jcp2}. Different works introduced approximative solutions of the SC equations beyond one-loop level in slit systems~\cite{hatlo,pre,japan}. Using a midpoint approximation that consists of replacing the local ionic self-energy by its value in the mid-pore, Yaroschuk solved the SC equations in cylindrical nanopores~\cite{yar1}. Then, within a restricted variational approach based on a uniform screening parameter in the pore, Buyukdagli {\it et al.}  improved the solution of Yaroschuk and showed that ion transport through nanoscale pores is characterized by an ionic conductor-insulator (CI) transition~\cite{PRL,jcp1}. Finally, Buyukdagli {\it et al.} recently developed a systematic numerical scheme for the exact solution of the SC equations in slit geometries~\cite{jcp2}. Comparisons with Monte-Carlo (MC) simulations showed that this SC scheme can handle surface dielectric discontinuities with a better accuracy than the DH approach.

Following these observations, we introduce in the present work an extended one-loop approach that is based on the truncation of the SC equations in such a way that the scheme reduces to the one-loop theory for dielectrically homogeneous systems, but considers surface polarization effects associated with low permittivity membranes in a self-consistent fashion. The latter point is indeed the most important advantage of our approach over numerical simulation schemes, since MC simulation techniques have been so far unable to deal with interfacial dielectric discontinuities in cylindrical pores. The article is organized as follows. We review in Section~\ref{scf} the physical framework of the SC formalism, and briefly introduce the extended one-loop theory whose detailed derivation is reported in Appendices~\ref{1lasym} and~\ref{ex1l}. Then, in Section~\ref{subsec1}, we make use of the present approach in order to analyze the MC simulation data of Refs.~\cite{mcasy,mcmix} for the partition of electrolyte mixtures in dielectrically homogeneous cylindrical pores. Finally, Section~\ref{subsec2} is devoted to dielectric discontinuity effects on the ionic selectivity of the pore, and our main results are summarized in the Conclusion part.

\section{Theory}

\subsection{Self-consistent formalism}
\label{scf}

\begin{figure}
\includegraphics[width=1.1\linewidth]{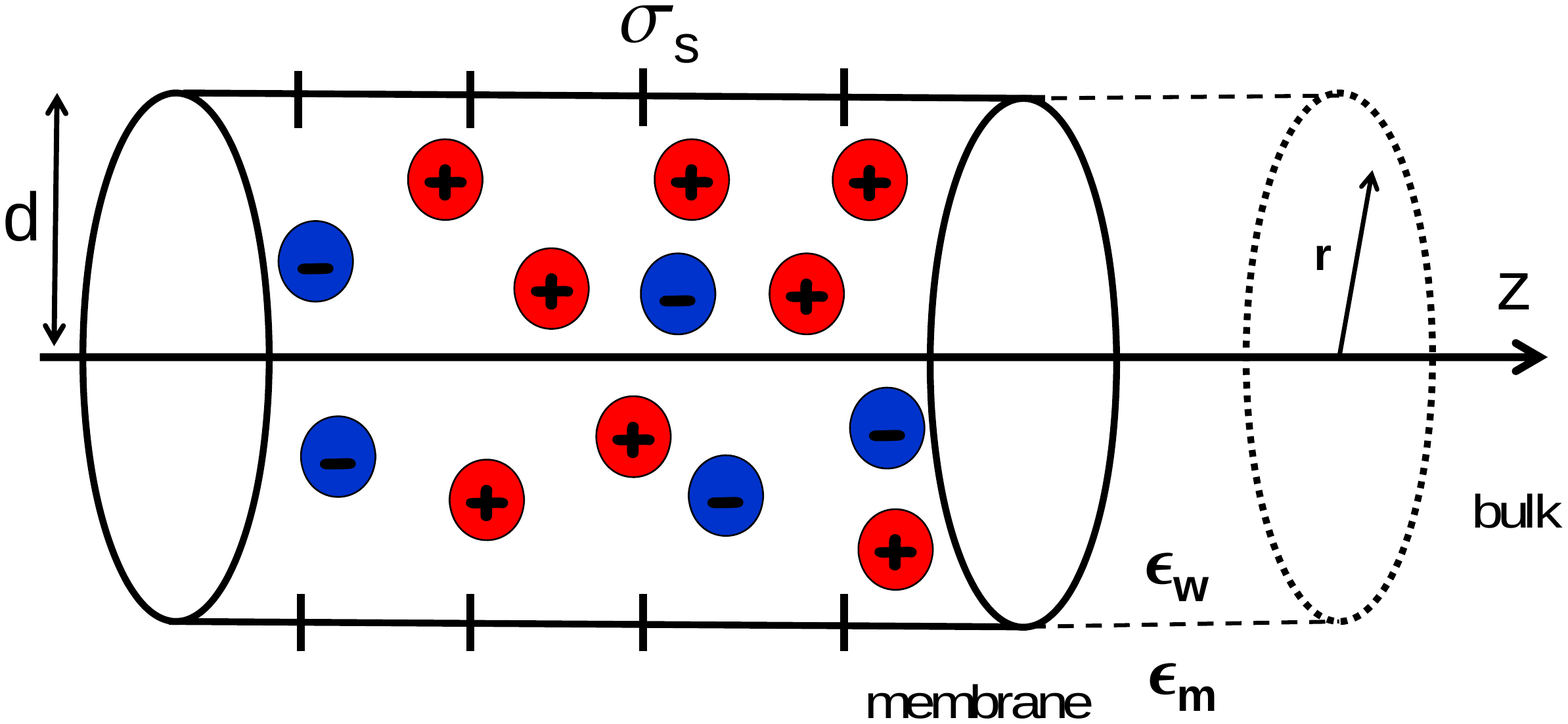}
\caption{(Color online) Schematic representation of the pore geometry. The pore radius is $d$, the net negative surface charge $-\sigma_s$, and the pore and membrane permittivities are respectively $\e_w=80$ and $\e_m\leq\e_w$.}
\label{fig0}
\end{figure}

We review in this part the SC formalism introduced in Refs.~\cite{SCrussian,netzvar}. The geometry of the membrane nanopore is depicted in Fig.~\ref{fig0}. It consists of a cylindrical pore of infinite length and radius $d$, and a negative wall charge distribution $\sigma(\br)=-\sigma_s\delta(r-d)$ with $\sigma_s>0$. The pore is in contact with an ion reservoir at the extremities, which is imposed by assuming the chemical equilibrium condition between the charges in the pore and the reservoir~\cite{jcp1}. Furthermore, the electrolyte is composed of $p$ ion species, with each species of reservoir concentration $\rho_{ib}$ and valency $q_i$. The number density of each ionic species is given by
\be\label{iden}
\rho_i(\br)=\rho_{ib}e^{-V_w(\br)}e^{-q_i\phi(\br)-\frac{q_i^2}{2}\delta v(\br)},
\ee
where the wall potential that imposes the ionic confinement inside the cylinder is introduced as $V_w(\br)=1$ if $0<r<d$, and $V_w(\br)=\infty$ for $r>d$. Moreover, the function $\phi(\br)$ in Eq.~(\ref{iden}) stands for the electrostatic potential generated by the fixed membrane charge at the pore wall or ionic charge excesses induced by charge separation effects. Then, the ionic self-energy is defined in the form of an equal point Green's function
\be\label{self}
\delta v(\br)\equiv \lim_{\br'\to\br}\left\{v(\br,\br')-v_c^b(\br-\br')+\ell_B\kappa_b\right\},
\ee
with the electrostatic Green's function $v(\br,\br')$  corresponding to the potential induced by an ion located at $r'$. In an ion free bulk reservoir, the latter reduces to the simple Coulomb potential $v_c^b(\br)=\ell_B/r$. Furthermore, the coefficient $\ell_B=7$ {\AA} in Eq.~(\ref{self}) stands for the Bjerrum length at ambient temperature $T=300$ $K$, and $\kappa_b^2=4\pi\ell_B\sum_i\rho_{ib}q_i^2$ is the DH screening parameter. Indeed, the self-energy introduced in Eq.~(\ref{self}) accounts for the spatial variations of the chemical potential of an ion resulting from the deformation of its screening cloud close to boundaries of the system (solvation energy) and its interaction with surface polarization charges (image charge forces).

The external potential $\phi(\br)$ and the Green's function $v(\br,\br')$ in Eqs.~(\ref{iden}) and~(\ref{self}) are solutions of the electrostatic self-consistent equations~\cite{SCrussian,netzvar},
\bea\label{eq1}
&&\nabla\e(\br)\nabla\phi(\br)+\frac{e^2}{k_BT}\sum_{i=1}^p\rho_i(\br)q_i=-\frac{e^2}{k_BT}\sigma(\br);\\
\label{eq2}
&&\nabla\e(\br)\nabla v(\br,\br')-\frac{e^2}{k_BT}\sum_{i=1}^p\rho_i(\br)q_i^2v(\br,\br')\nonumber\\
&&=-\frac{e^2}{k_BT}\delta(\br-\br'),
\eea
where the dielectric permittivity function accounting for the variations of the background permittivity at the boundaries of the cylindrical pore is given by $\e(\br)=\e_w\theta(d-r)+\e_m\theta(r-d)$, with $\e_w\simeq 80$ and $\e_m\leq \e_w$ the water and membrane permittivites, respectively. Moreover, $e$ stands for the elementary charge, and $k_B$ the Boltzmann constant. One notes that Eq.~(\ref{eq1}) is an extended Poisson-Boltzmann (PB) equation for the electrostatic potential $\phi(\br)$. Furthermore,  Eq.~(\ref{eq2}) corresponds to a generalized Debye-Huckel (DH) equation for the Green's function $v(\br,\br')$. The difference from the usual PB and DH equations is respectively due to the non-uniform ion density $\rho_i(\br)$ in Eq.~(\ref{eq2}), and the presence of the ionic self-energy term $\delta v(\br)$ in Eq.~(\ref{iden}) that introduces in Eq.~(\ref{eq1}) an inhomogeneous screening of the external potential. \textcolor{black}{We also note that the bulk limit of Eqs.~(\ref{eq1})-(\ref{eq2}) was investigated in Ref.~\cite{pre}. Indeed, it was shown that for monovalent ions with concentration $\rho_{ib}<2.0$ M,  the solution of the variational equations yields the DH limiting law, whereas for higher concentrations, the electrolytic free energy becomes unstable. Although this instability can be avoided with an ultraviolet cut-off, the regularization is unnecessary since we have previously showed that the instability regime lies well beyond the ion density range where the SC formalism is quantitatively accurate~\cite{jcp2}.} 

To characterize the correlation effects \textcolor{black} {induced by the confinement} on the charge selectivity of the pore, it is helpful to identify the key system parameters. To this end, we consider a dielectrically homogeneous pore $\e_m=\e_w$ confining a symmetric electrolyte composed of two ion species of valencies $q_{\pm}=\pm q$, with $q>0$. Rescaling the distances in Eqs.~(\ref{eq1})-(\ref{eq2}) according to $\brr=\kappa_b r$, and renormalizing the external potential and the Green's function as $\psi(\bbr)=q\phi(\bbr)$ and $u(\bbr,\bbr')=q^2v(\bbr,\bbr')$, the SC equations take the form
\bea\label{eq3}
&&\nabla^2\psi(\bbr)-\theta(\bd-\brr)e^{-\frac{1}{2}\delta u(\bbr)}\sinh\psi(\bbr)=\frac{2}{s}\delta(\brr-\bd);\\
\label{eq4}
&&\nabla^2u(\bbr,\bbr')-\theta(\bd-\brr)e^{-\frac{1}{2}\delta u(\bbr)}\cosh\psi(\bbr)u(\bbr,\bbr')\nonumber\\
&&=-4\pi\Gamma\delta(\bbr-\bbr').
\eea
The rescaled equations~(\ref{eq3})-(\ref{eq4}) show that for dielectrically homogeneous pores and symmetric electrolytes, the ion density in the pore is solely characterized by three parameters. These are the reduced pore radius $\bd=\kappa_b d$, the coupling parameter $\Gamma=q^2\kappa_b\ell_B$ characterizing the strength of the electrostatic potential fluctuations around the MF potential, and the parameter $s=\kappa_b\mu$, where $\mu=1/(2\pi q\ell_B\sigma_s)$ stands for the Gouy-Chapman length. Namely, the parameter $s$ measures the ratio of the counterion layer thickness $\mu$ at the charged pore wall to the screening cloud radius $\kappa_b^{-1}$ around a central ion in the bulk reservoir. In Section~\ref{subsec1} on electrolytes in dielectrically homogeneous pores, it will be shown that electrostatic correlation effects are characterized by an interpolation between the parameter regime $s\ll1$ corresponding to a compact interfacial counterion layer and the regime $s\gg1$ associated with a diffuse counterion partition close to the pore wall.

In the present work, we will generalize to dielectrically discontinuous pores the one-loop approach developed in Refs.~\cite{netzcoun,1loop,jcp2} for dielectrically uniform systems. Our strategy consists in truncating the SC Eqs.~(\ref{eq1})-(\ref{eq2}) in such a way that the extended scheme reduces to the one-loop theory for dielectrically uniform pores with $\e_m=\e_w$. The details of the derivation of the extended approach that we summarize in Section~\ref{trunc} are reported in Appendix~\ref{ex1l}.

\subsection{Truncating SC equations}
\label{trunc}

The one-loop theory of a symmetric electrolyte in contact with a dielectrically continuous charged plane was shown in Ref.~\cite{jcp2} to follow from the linearization of the SC Eqs.~(\ref{eq1})-(\ref{eq2}) in terms of the electrostatic Green's function $v(\br,\br')$. The extension of the one-loop approach to a general electrolyte mixture is presented in Appendix~\ref{1lasym}. In the present work, we also introduce a generalization of the one-loop approach to dielectrically discontinuous systems where the singular nature of the self-energy on the pore wall does not allow a simple Taylor expansion of Eq.~(\ref{iden})~\cite{David,jcp2}. Our strategy that we detail in Appendix~\ref{ex1l} consists in treating in Eqs.~(\ref{eq1})-(\ref{eq2}) the part of the Green's function induced by the dielectric discontinuity self-consistently, and expanding the rest resulting from solvation forces perturbatively.

To derive the truncated equations, we formally split the self-energy into a solvation and an image charge contribution as $\delta v(\br)=\delta v^{(im)}(\br)+\delta v^{(s)}(\br)$, where the potential $\delta v^{(im)}(\br)$ accounts for the interaction of ions with the interfacial polarization charges, and the potential $\delta v^{(s)}(\br)$ incorporates the solvation energy induced by the deformation of the ionic cloud in the pore. Then we insert this expansion into Eqs.~(\ref{eq1})-(\ref{eq2}), and linearize the latter in terms of the solvation energy $\delta v^{(s)}(\br)$. This truncation results in an expansion of the external potential in the form $\phi(\br)=\phi_0(\br)+\phi_1(\br)$, where the component $\phi_0(\br)$ of the potential and the Green's function $v(\br,\br')$ satisfy the differential equations
\bea\label{eqt1}
&&\nabla\e(\br)\nabla\phi_0(\br)+\frac{e^2}{k_BT}\sum_{i=1}^pq_in_i(\br)=-\frac{e^2}{k_BT}\sigma(\br);\\
\label{eqt2}
&&\nabla\e(\br)\nabla v(\br,\br')-\frac{e^2}{k_BT}\sum_{i=1}^pq_i^2n_i(\br)v(\br,\br')\nonumber\\
&&=-\frac{e^2}{k_BT}\delta(\br-\br'),
\eea
and the correction term to the external potential is given by
\be\label{eqt3}
\phi_1(\br)=\int\mathrm{d}\br'v(\br,\br')\delta\sigma(\br'),
\ee
with the auxiliary number and excess charge densities
\bea\label{aux1}
n_i(\br)&=&\rho_{ib}e^{-V_w(\br)-q_i\phi_0(\br)}e^{-\frac{q_i^2}{2}\delta v^{(im)}(\br)};\\
\label{aux2}
\delta\sigma(\br)&=&-\frac{1}{2}\sum_{i=1}^pq_i^3n_i(\br)\delta v^{(s)}(\br).
\eea
One sees that Eq.~(\ref{eqt1}) is a modified PB equation that accounts for the variations of the external potential by the image charge forces. Furthermore, Eqs.~(\ref{eqt3}) and~(\ref{aux2}) indicate that the term $\phi_1(\br)$ brings corrections from solvation forces. The explicit form of the potentials $\delta v^{(s)}(\br)$ and $\delta v^{(im)}(\br)$ as well as the iterative solution scheme of Eqs.~(\ref{eqt1})-(\ref{eqt3}) is explained in Appendix~\ref{ex1l}. We finally note that in the limit $\e_m=\e_w$, Eqs.~(\ref{eqt1})-(\ref{eqt3}) reduce to the one-loop equations derived in Appendix~\ref{1lasym}.

\section{Results}

\subsection{Dielectrically homogeneous pores}
\label{subsec1}

We characterize in this section ionic correlation effects on the selectivity of dielectrically homogeneous pores ($\e_m=\e_w$) confining electrolytes of general mixture. The theoretical ion density results will be compared with MC simulation data from Refs.~\cite{mcasy,mcmix} in order to establish the quantitative accuracy of the present theory. The transparency of our theoretical scheme will also allow to probe in detail the underlying physics behind the simulation results. Section~\ref{subsec1I} is devoted to correlation effects on the partition of asymmetric electrolytes composed of two species, and we will consider in Section~\ref{subsec1II} the more complicated case of mixed electrolytes.

\subsubsection{Asymmetric electrolytes}
\label{subsec1I}

\begin{figure}
\includegraphics[width=1.1\linewidth]{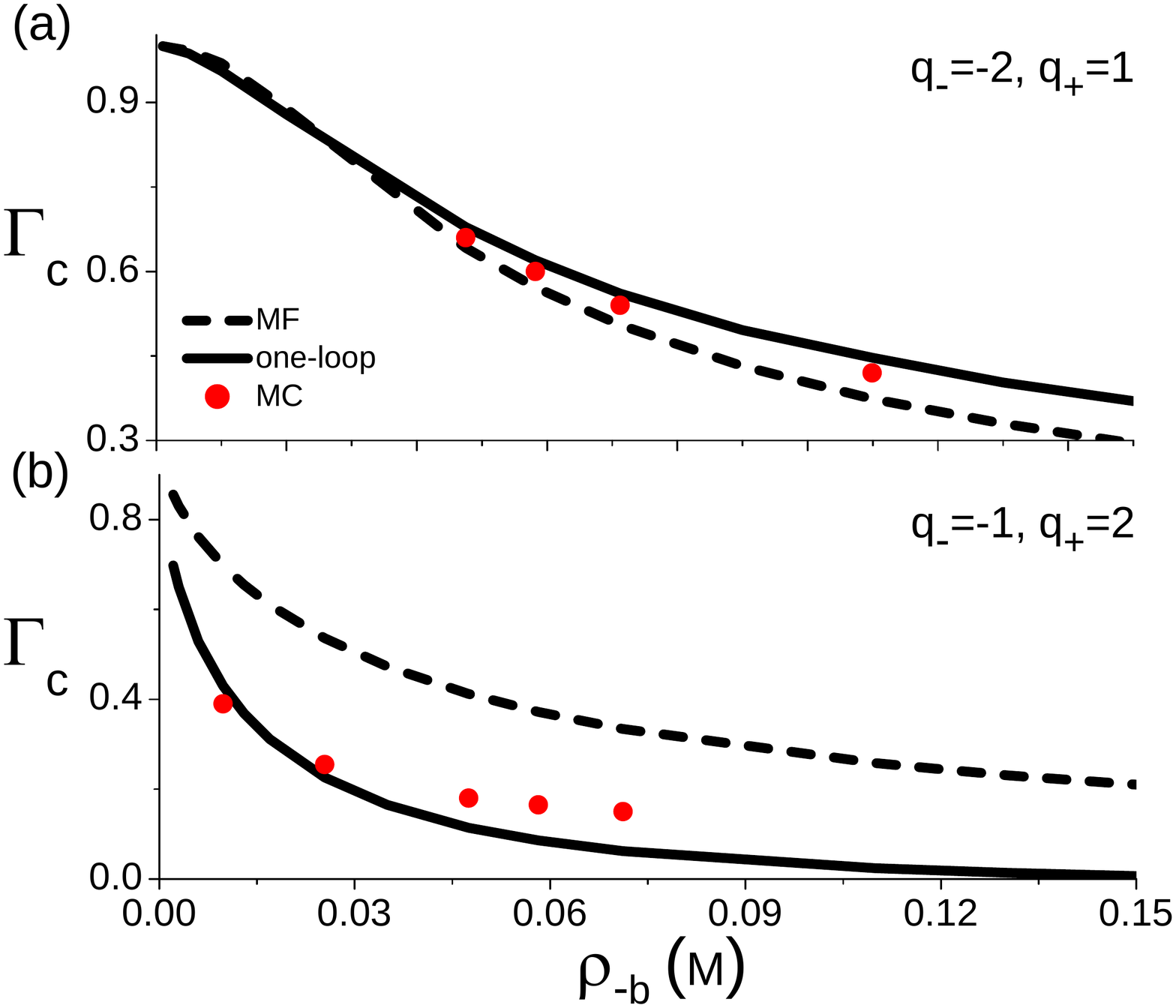}
\caption{(Color online) Coion rejection rates against the anion density for (a) divalent anions $q_-=-2$ and monovalent counterions $q_+=1$, and (b) monovalent anions $q_-=-1$ and divalent counterions $q_+=2$. The pore charge is $\sigma_s=0.0356$ $\mathrm{C/m}^2$, pore radius $d=4$ nm, and the membrane permittivity $\e_m=\e_w$. The dashed and solid black curves are respectively the MF and one-loop results, and the MC data in (a) and (b) denoted by the red circles are respectively taken from Tables III and IV of Ref~\cite{mcasy}.}
\label{fig1}
\end{figure}

\begin{figure}
\includegraphics[width=1.0\linewidth]{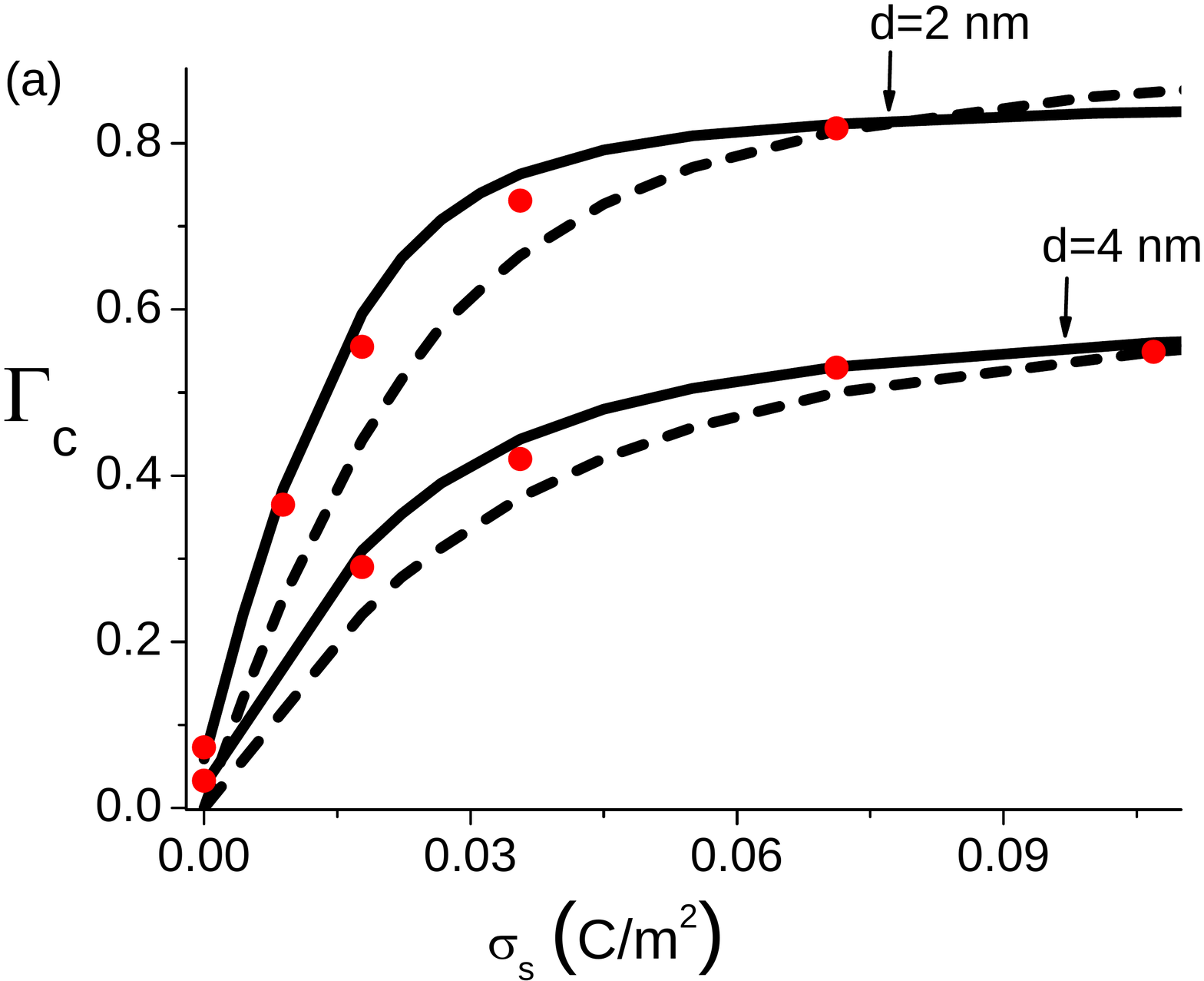}
\includegraphics[width=1.0\linewidth]{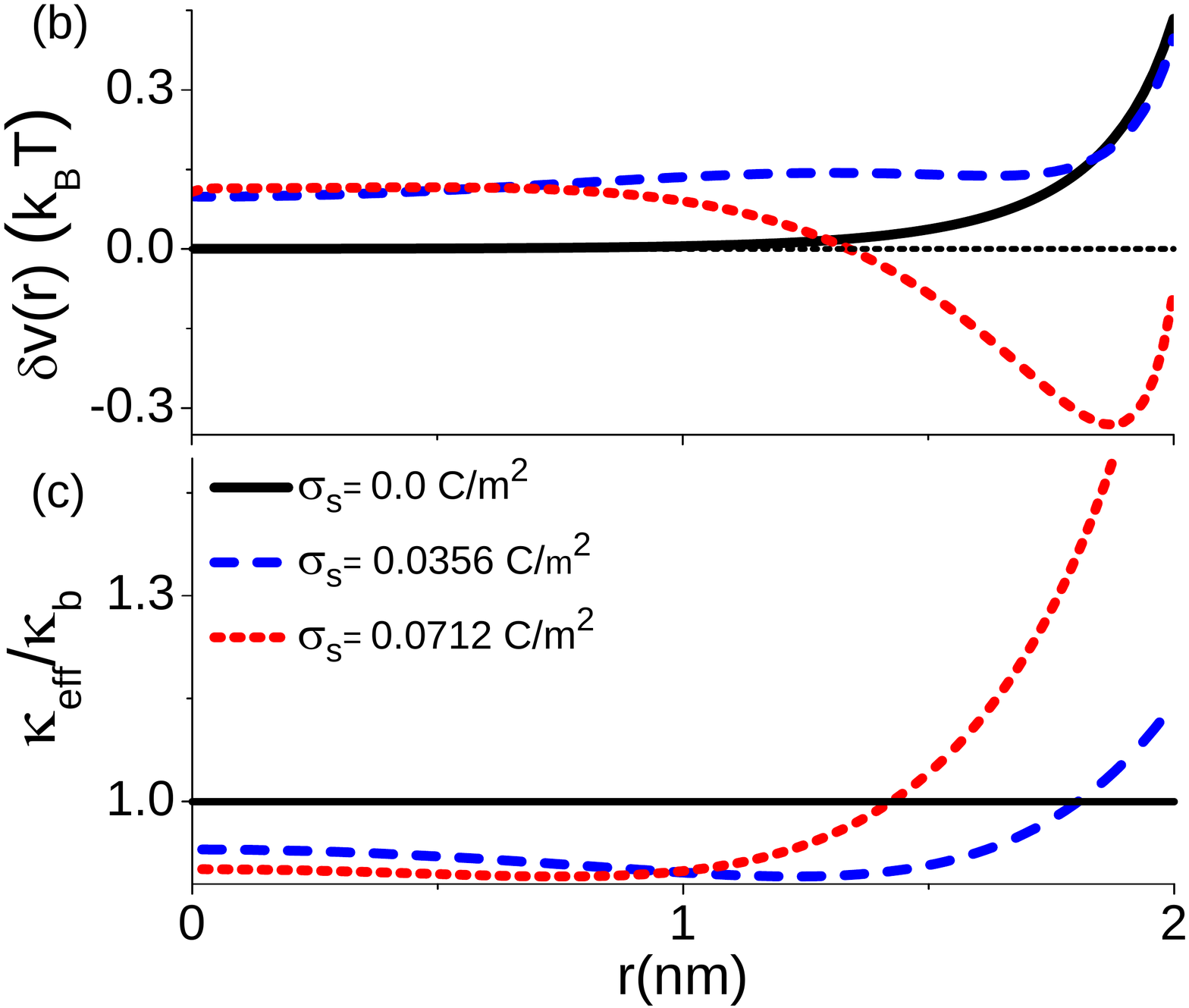}
\caption{(Color online) (a) Coion rejection rates against the pore charge for divalent anions $q_-=-2$, monovalent counterions $q_+=1$, with the bulk anion density $\rho_{-b}=0.1099$ M, and membrane permittivity $\e_m=\e_w$. The curves and symbols have the same meaning as in Fig.~\ref{fig1}. MC data were taken from Table III of Ref~\cite{mcasy}. (b) Ionic self energy and (c) the effective screening parameter of Eq.~(\ref{kapef}) for the same system parameters as in (a) and the surface charges given in the legend.}
\label{fig2}
\end{figure}

In this part, we focus on the MC simulation results of Ref.~\cite{mcasy} for the partition of asymmetric solutions and scrutinize in detail the underlying electrostatic interactions. The reservoir contains an asymmetric electrolyte composed of two ion species. We first consider in Fig.~\ref{fig1}(a) the case of divalent coions and monovalent counterions ($q_-=-2$ and $q_+=1$), with the pore radius $d=4$ nm and charge $\sigma_s=0.0356$ $\mathrm{C/m}^2$.  The figure displays the rejection rates defined as $\Gamma_c=1-k_-$, where the ionic partition function corresponds to the pore averaged ion density,
\be\label{defpart}
k_i=\frac{2}{d^2\rho_b}\int_0^d\mathrm{d}rr\rho_i(r),
\ee
with the local ion densities given by (see Appendix~\ref{1lasym})
\be\label{den1l}
\rho_i(r)=\rho_{ib}e^{-q_i\phi_0(r)}\left[1-q_i\phi_1(r)-\frac{q_i^2}{2} \delta v(r)\right].
\ee
It is seen that over the bulk density regime $\rho_{-b}>0.03$ M, the MC data (red dots) and the one-loop result (solid black curves) in Fig.~\ref{fig1}(a) predict a slightly higher rejection rate than the MF theory (dashed black curve). The same effect is also illustrated in Fig.~\ref{fig2}(a) that displays the coion rejection rates against the surface charge at fixed bulk concentration $\rho_{-b}=0.1099$ M, and for pore radii $r=2$ nm and $4$ nm. To understand the underestimation of the rejection rates by the MF theory, one should note that for monovalent counterions, the surface charge densities $\sigma_s<0.068$ $\mathrm{C/m}^2$ in Fig.~\ref{fig2}(a) correspond to the parameter regime $s>1$  characterized by a diffuse counterion layer next to the pore wall. In this regime where the ion free membrane is responsible for a charge screening deficiency in the vicinity of the charged wall, ions feel a repulsion from the wall towards the mid-pore area where they are more efficiently screened. Thus, correlations decrease the MF density of both anions and cations in the pore. The positive ionic self-energy embodying this repulsive force $\delta v(r)$ is reported in Fig.~\ref{fig2}(b) (solid black curve).

An interesting point to be noted in Fig.~\ref{fig2}(b) is the increase of the ionic self-energy in the mid-pore with an increase of the surface charge from $\sigma_s=0$ $\mathrm{C/m}^2$ to $0.0356$ $\mathrm{C/m}^2$. The positive energy barrier in the mid-pore area that appears at finite charges is indeed the second correlation effect behind the deviation of the MF curves from the one-loop and MC results in Fig.~\ref{fig2}(a). This peculiarity can be understood in terms of the effective screening parameter
\be\label{kapef}
\kappa^2_{eff}(r)=4\pi\ell_B\sum_{i=1}^pq_i^2n_i(r)
\ee
that determines the local screening of the one-loop potential in Eq.~(\ref{eq7}), with the function $n(r)$ defined in Eq.~(\ref{aux1}). In Fig.~\ref{fig2}(c), it is seen that while moving from the charged wall towards the mid-pore, the local screening parameter drops below the bulk one. This originates from the exclusion of divalent coions by the surface charge, which dominates the monovalent counterion excess in the pore. The corresponding charge screening deficiency with respect to the bulk reservoir results in an electrostatic energy barrier for ionic penetration. Then, by further increasing the surface charge from $\sigma_s=0.0356$ $\mathrm{C/m}^2$ to $0.0712$ $\mathrm{C/m}^2$, one gets into the second regime $s<1$ where the counterions form a dense layer in the vicinity of the charged wall. The resulting interfacial ionic screening excess with respect to the bulk reservoir (see Fig.~\ref{fig2}(c)) leads to a negative ionic self-energy close to the pore wall, although the mid pore area is still characterized by the strong coion deficiency resulting in the positive branch of the self-energy (see Fig.~\ref{fig2}(b)). As a result, the one-loop rejection rates in Fig.~\ref{fig2}(a) drop below the MF result for large pore charges. One can also notice in Fig.~\ref{fig2}(a) the good agreement between the one-loop theory and the MC data, which is remarkable if one considers the large coupling parameter $\Gamma\simeq 5.2$ for divalent anions at the bulk density $\rho_{-b}=0.1099$ M.

\begin{figure}
\includegraphics[width=1.0\linewidth]{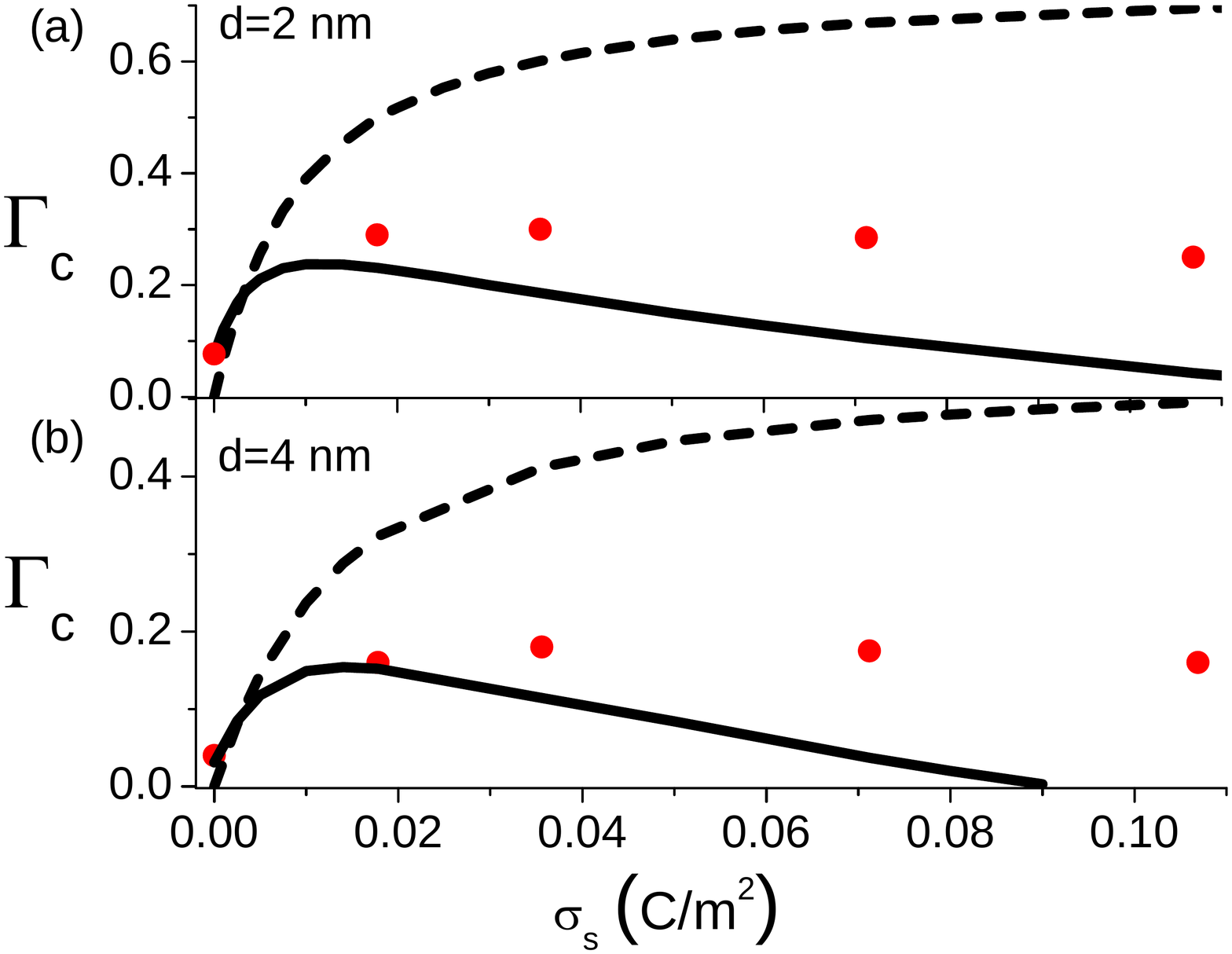}
\includegraphics[width=1.0\linewidth]{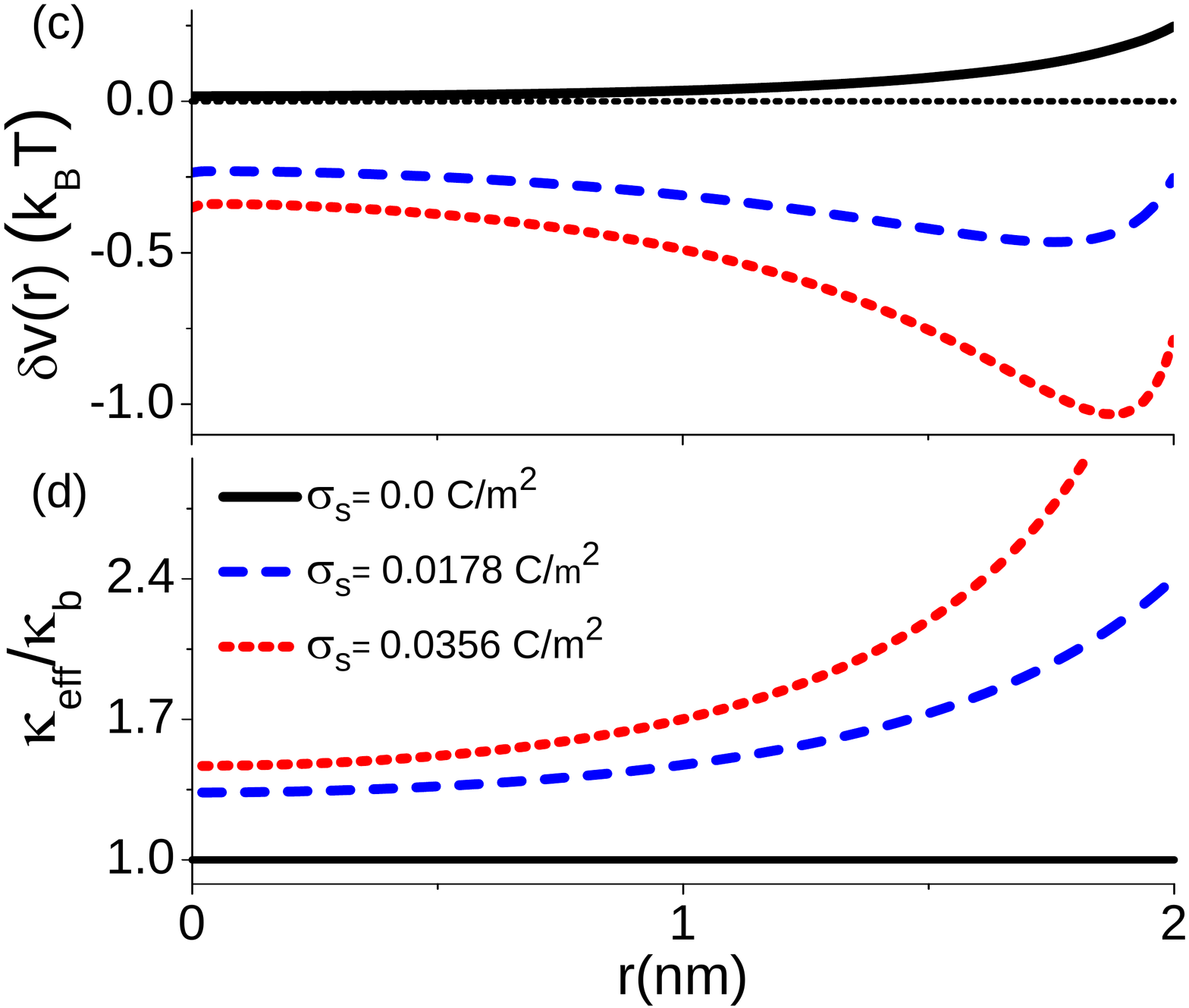}
\caption{(Color online) Coion rejection rates for pore radius (a) $d=2$ nm and (b) $d=4$ nm against the pore charge for monovalent coions $q_-=-1$, divalent counterions $q_+=2$, with the bulk anion density $\rho_{-b}=0.0475$ M, and the membrane permittivity $\e_m=\e_w$. The curves and symbols have the same meaning as in Fig.~\ref{fig1}. MC data were taken from Table IV of Ref~\cite{mcasy}. (c) Ionic self energy and (d) effective screening parameter Eq.~(\ref{kapef}) for the same system parameters as in (a) and the surface charges given in the legend.}
\label{fig3}
\end{figure}

\begin{figure}
\includegraphics[width=1.\linewidth]{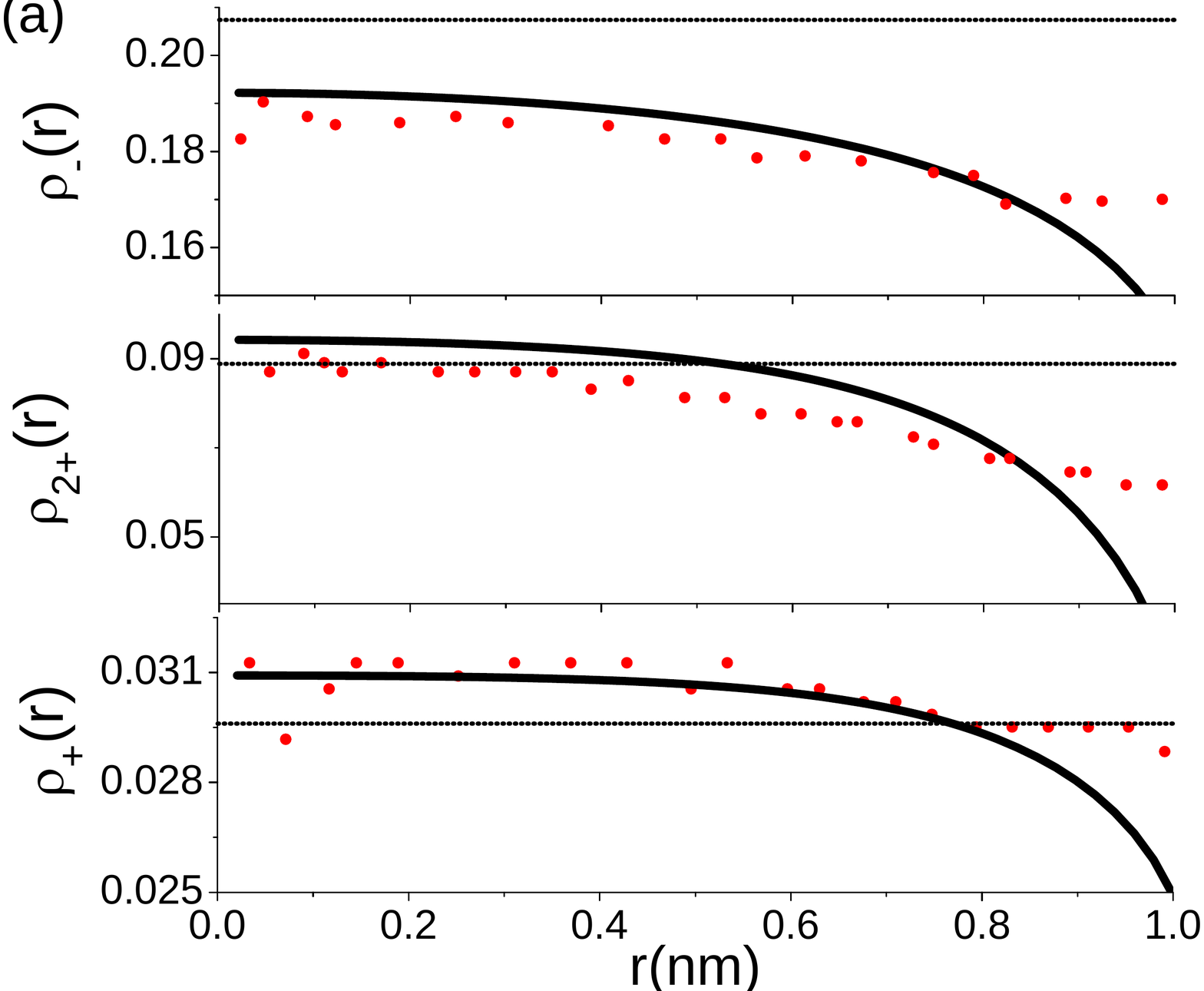}
\includegraphics[width=1.\linewidth]{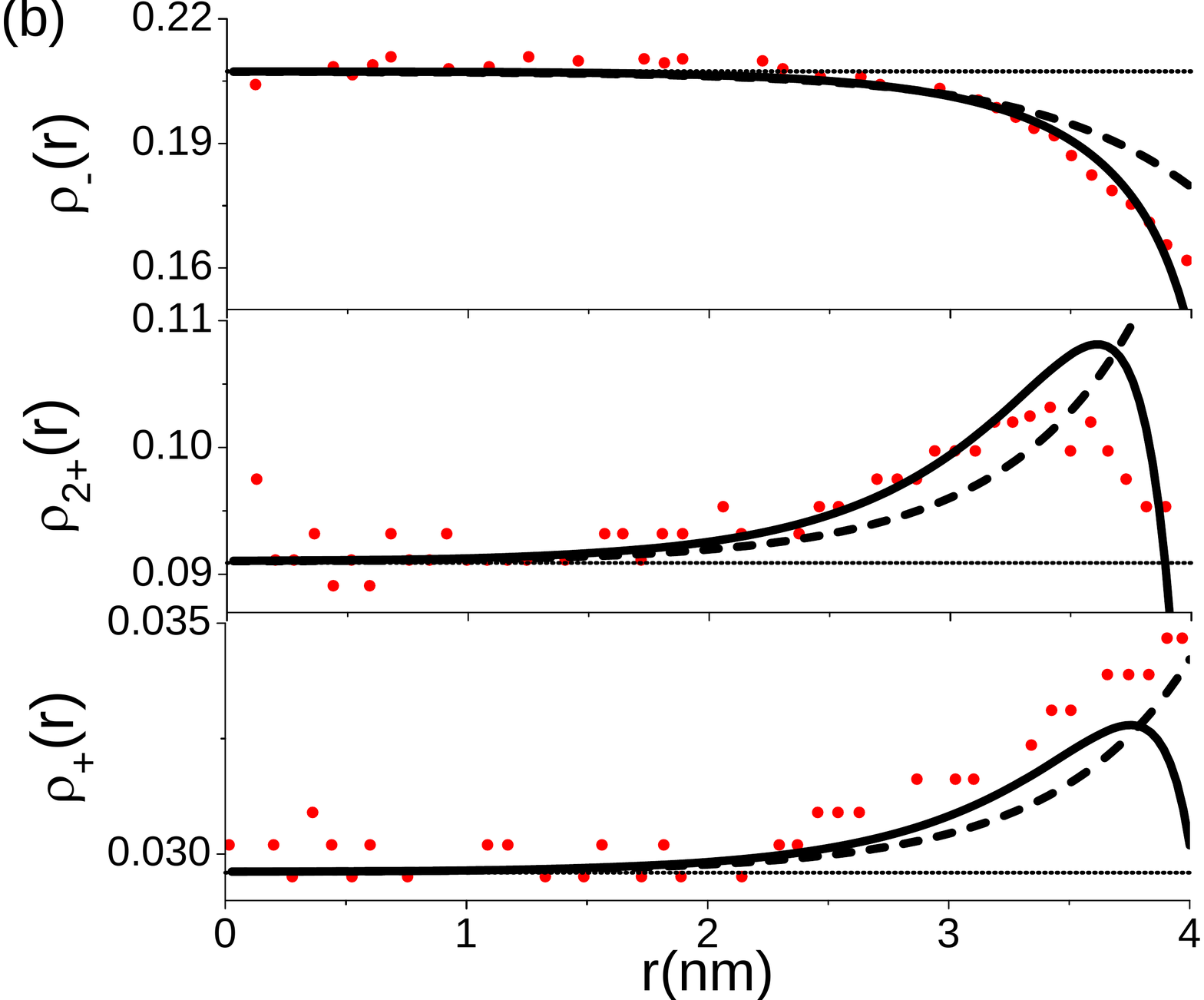}
\includegraphics[width=1.\linewidth]{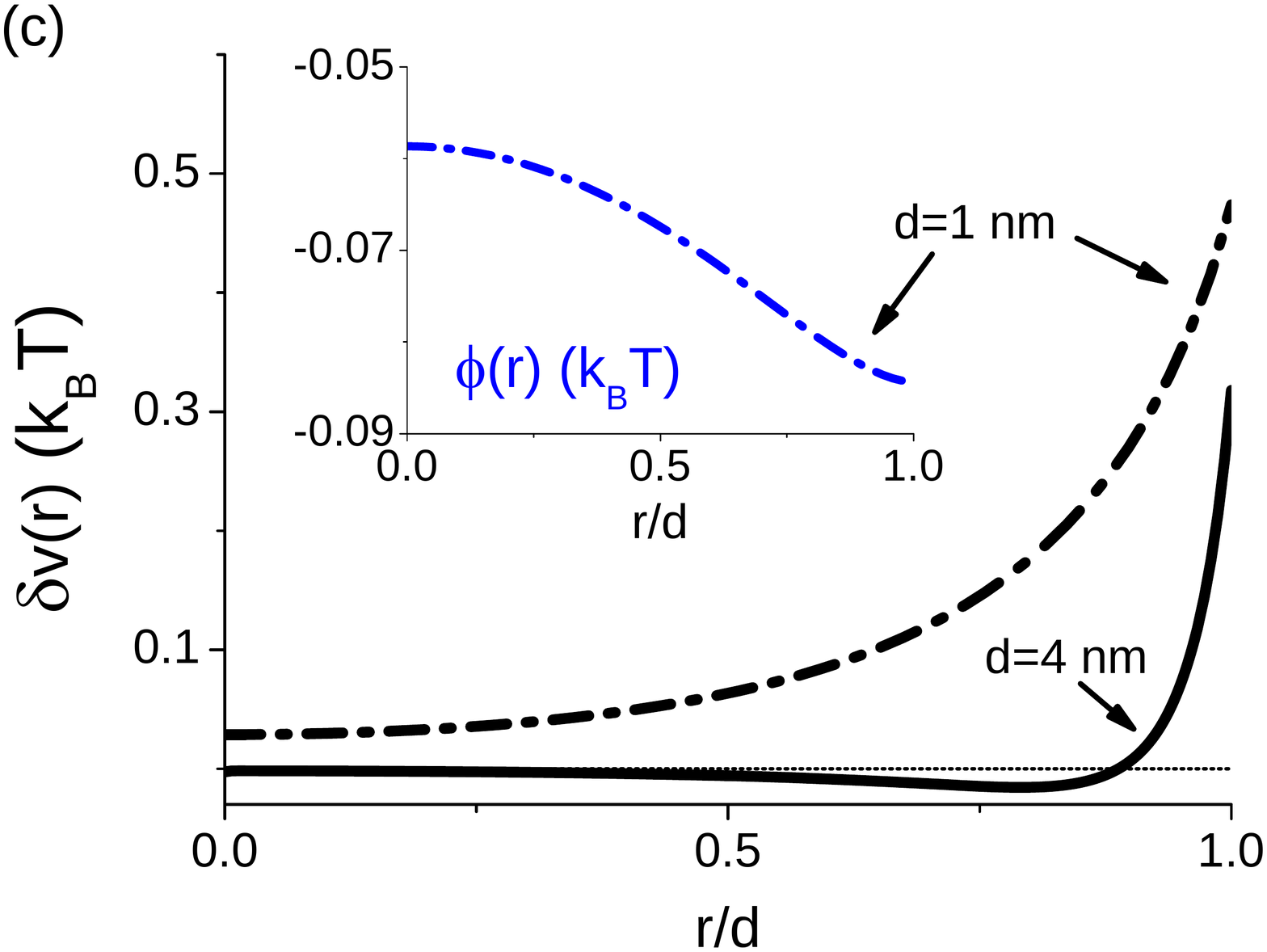}
\caption{(Color online) Ion density profiles for the electrolyte mixture $\mathrm{CaCl}_2$ and $\mathrm{KCl}$ in the nanopore with (a) vanishing surface charge $\sigma_s=0$ $\mathrm{C/m}^2$ and radius $d=1$ nm, and (b) finite surface charge $\sigma_s=0.00445$ $\mathrm{C/m}^2$ and radius $d=4$ nm.  Bulk ion densities are $\rho_{-b}=0.2074$ M, $\rho_{2+b}=0.0889$ M and $\rho_{+b}=0.0296$ M, and membrane permittivity is $\e_m=\e_w$. The curves and symbols have the same meaning as in Fig.~\ref{fig1}. MC data in (a) and (b) were respectively taken from Figs. 4 and 12 of Ref~\cite{mcmix}. (c) External potential (inset) and ionic self energy (main plot) for the same system parameters as in (a) (dashed-dotted curves) and (b) (solid curve).}
\label{fig4}
\end{figure}
We next consider an asymmetric electrolyte composed of monovalent coions and divalent counterions. Fig.~\ref{fig1}(b) displays for this case the anion rejection rates against the bulk density at the pore charge $\sigma_s=0.0356$ $\mathrm{C/m}^2$. The inspection of the figure shows that the one-loop result remains very close to the MC data, significantly improving over the MF result. Then, it is seen that the latter now overestimates the MC and one-loop rejection curves. This effect is also illustrated in Figs.~\ref{fig3}(a) and (b) over the whole surface charge regime $\sigma_s\leq0.11$ $\mathrm{C/m}^2$. To understand this peculiarity, one should note that for divalent counterions, the pore is characterized by a significant counterion excess even for weak surface charges. The resulting screening excess $\kappa_{eff}(r)>\kappa_b$ displayed in Fig.~\ref{fig3}(d) is shown in Fig.~\ref{fig3}(c) to yield a significantly negative ionic self-energy on the order of the thermal energy $k_BT$, which is monotonically lowered with increasing surface charge. As a result of this pure correlation effect, the coion rejection rates in Fig.~\ref{fig3}(a) obtained from MC and one-loop approaches are seen to reach a peak at a characteristic surface charge, and to decrease for higher pore charges where the screening excess in the pore takes over the Donnan exclusion. \textcolor{black}{As noted in Ref.~\cite{mcasy},}  this feature may be the precursor of the negative rejection of electrolytes with multivalent counterions observed in ion rejection experiments with porous glass~\cite{exp0}. Thus, the enhancement of coion densities with increasing surface charge clearly deserves further investigation, and we will reconsider the effect in Section~\ref{enh} in further detail.

Finally, in Figs.~\ref{fig3}(a) and (b), the one-loop theory is seen to exhibit a quantitative agreement with MC data exclusively at low surface charges. Namely, the turning point where the coion density starts to increase with the surface charge is seen to be underestimated by the one-loop approach, and the deviation increases beyond this value. Although the discrepancy should be mainly due to electrostatic correlation effects, we also expect the hard-core (HC) interactions absent in the one-loop theory to be partly responsible for the disagreement.

\subsubsection{Electrolyte mixtures}
\label{subsec1II}

We consider in this part the simulation results of Ref.~\cite{mcmix} for the partition of an electrolyte mixture composed of a symmetric and an asymmetric electrolyte, such as $\mathrm{CaCl}_2$ and $\mathrm{KCl}$. Fig.~\ref{fig4}(a) compares the one-loop ion densities with the MC data in a narrow pore with radius $d=1$ nm and vanishing surface charge $\sigma_s=0$ $\mathrm{C/m}^2$. The ion densities in the reservoir are $\rho_{-b}=0.2074$ M, $\rho_{2+b}=0.0889$ M and $\rho_{+b}=0.0296$ M, which are marked in the plots by the dotted horizontal curves. First of all, it is seen that despite the large coion density and strong confinement, the theory agrees well with MC simulations within the numerical uncertainties of the data, except in the vicinity of the pore wall where the theory predicts a stronger ion depletion. \textcolor{black}{This deviation may be due to the fact that our formalism neglects the interfacial ion-free layer associated with the finite ionic radius. This layer could be easily incorporated into the Green's function computed in Appendix~\ref{DHsol} as it was done in Refs~\cite{lev2,lev3} in planar geometry and in Ref~\cite{jcp1} for cylinders, but we leave this improvement for a future work for the sake of simplicity. We also note in passing that for ions in a neutral pore, the MF theory yields the ideal gas behavior $\rho_i(r)=\rho_{ib}$, which is denoted in Figs.~\ref{fig4} (a) and (b) by the horizontal lines.}

Two features to be noted in Fig.~\ref{fig4}(a) are an overall depletion of $\mathrm{Cl}^-$ and $\mathrm{Ca}^{2+}$ ions, and a weak pore excess of $\mathrm{K}^+$ ions. To explain these points, we report in Fig.~\ref{fig4}(c) the ionic self-energy (dashed-dotted curve in the main plot) and the external potential (inset). It is seen that the deformation of the screening cloud around the ions results in a positive self-energy, i.e. an electrostatic energy barrier depleting the monovalent anions and divalent cations from the pore. Furthermore, as a consequence of the stronger exclusion of divalent cations with respect to monovalent anions, the bulk electroneutrality is locally perturbated in the pore. The corresponding charge separation effect leads in turn to a weakly negative external potential (see the inset of Fig.~\ref{fig4}(c)), resulting in a $\mathrm{K}^+$ excess in the mid-pore area.

We next consider in Fig.~\ref{fig4}(b) a larger pore of radius $d=4$ nm and surface charge $\sigma_s=0.00445$ $\mathrm{C/m}^2$. The bulk ion densities are the same as in Fig.~\ref{fig4}(a). The theoretical density curves are again seen to yield a reasonable agreement with MC simulation data that exhibit large uncertainties. In this case, the counterion excess induced by the surface charge at the pore wall (see Fig.~\ref{fig4}(b)) is shown in Fig.~\ref{fig4}(c) to yield a weakly negative ionic self-energy, except in the vicinity of the pore wall where the screening defficiency takes over and results in a positive branch of the self-energy. In Fig.~\ref{fig4}(b), one sees that the repulsive branch leads to an ionic depletion layer close to the wall, and its competition with the surface charge results in a concentration peak for counterions. Moreover, in the same figure, the attractive branch of the self-energy is shown to slightly increase the MF prediction of cation densities. In the Section~\ref{subsec2}, we will incorporate into this picture surface polarization charges resulting from the dielectric discontinuity between the pore and the low permittivity membrane.

\subsection{Dielectrically inhomogeneous pores}
\label{subsec2}
\begin{figure}
\includegraphics[width=1.1\linewidth]{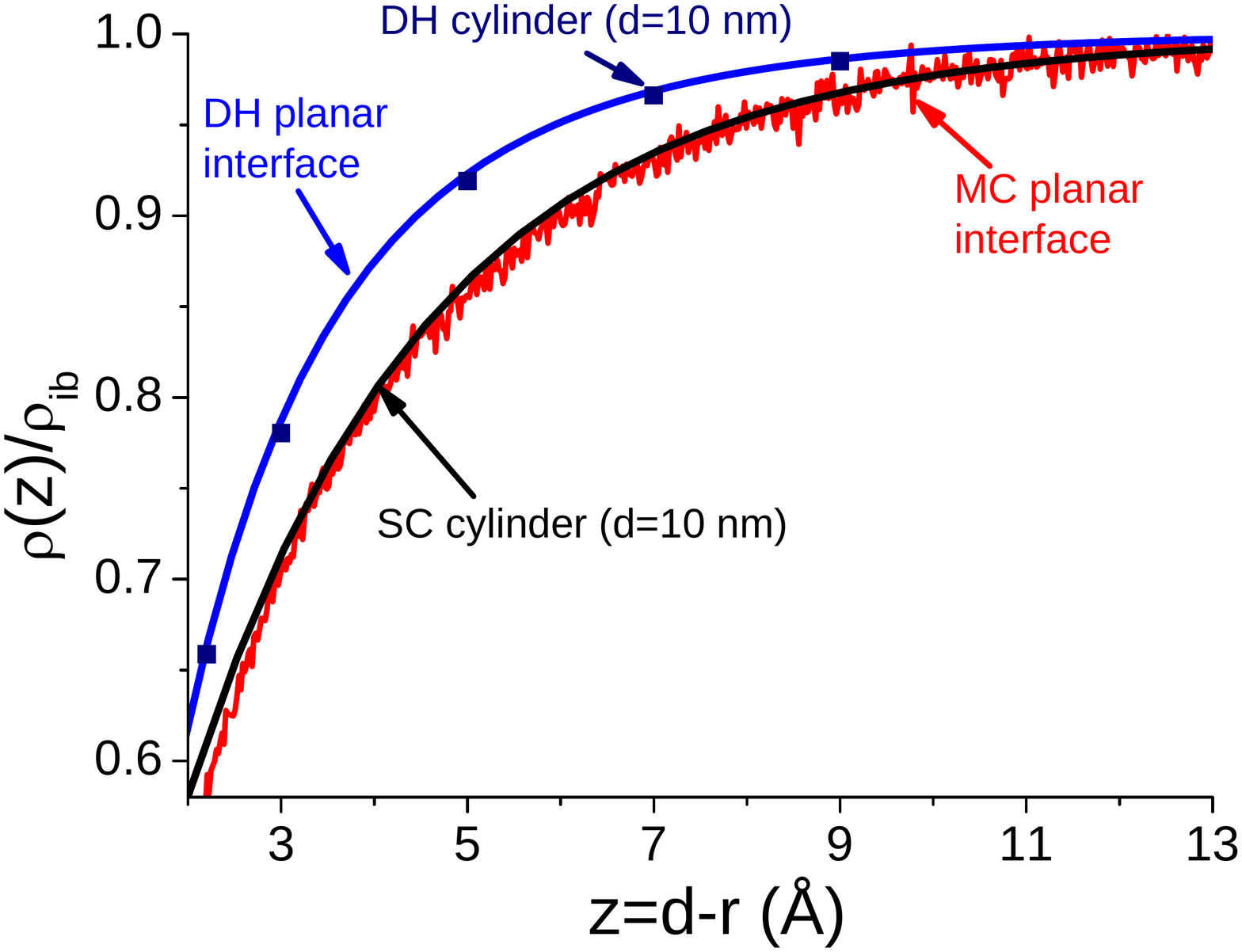}
\caption{(Color online) Ion density against the distance from the neutral pore wall for a symmetric electrolyte of two monovalent ion species. The solid red and blue curves are respectively MC simulation data from Ref.~\cite{jcp2} and the DH density for ions at a planar dielectric interface, and the square symbols and solid black curve respectively mark for a cylinder of radius $d=10$ nm the ion density profile from the DH approximation and the SC scheme introduced in Appendix~\ref{ex1l}. The bulk salt concentration is $\rho_{ib}=0.2$ M, and the permittivities are $\e_m=1$ and $\e_w=80$.}
\label{fig5}
\end{figure}

We consider in this part ionic partitions in cylindrical nanopores separating the electrolyte from a membrane medium with a low dielectric permittivity $\e_m\leq\e_w=80$. The SC solution of the electrostatic closure equations~(\ref{eqt1})-(\ref{eqt3}) for dielectrically inhomogeneous pores is explained in Appendix~\ref{ex1l}. For this more general case, the ion density is given by
\be\label{denSC}
\rho_i(r)=\rho_{ib}\;e^{-q_i\phi_0(r)-\frac{q_i^2}{2}\delta v_{im}(r)}\left[1-q_i\phi_1(r)-\frac{q_i^2}{2} \delta v_s(r)\right],
\ee
where the external potential $\phi_0(r)$ is the solution of the MPB equation~(\ref{sp3}), the correction to the external potential $\phi_1(r)$ is obtained from Eq.~(\ref{spl10}), and the solvation potential $\delta v_s(r)$ and the image-charge energy $\delta v_{im}(r)$ resulting from the dielectric discontinuity at $r=d$ are given by Eq.~(\ref{spl15}).

\subsubsection{Comparison with MC simulations}

We note that the present SC scheme that treats solvation forces in a perturbative fashion differs from the approach developed in Ref.~\cite{jcp2} for ions at dielectric planes. Although there are no available MC simulation data for ion densities in dielectrically discontinuous cylinders, we can still test the quantitative validity of the present theory for cylinders with a large radius by comparison with MC simulations at planar dielectric interfaces. To this end, we compare in Fig.~\ref{fig5} the MC data from Ref.~\cite{jcp2} for a symmetric electrolyte at a neutral dielectric plane (red curve), with the predictions of the present scheme for a cylinder of radius $d=10$ nm (black curve). The bulk ion concentration is $\rho_{ib}=0.2$ M, and the membrane permittivity $\e_m=1$. In order to illustrate the corrections beyond the WC regime, we also report the DH density profiles for a planar interface~\cite{jcp2} (solid blue curve) and a cylinder of radius $d=10$ nm (square symbols). The DH density profile for the cylinder is computed from the relation $\rho_i(r)=\rho_{ib}\exp\left[-q^2\delta v_{DH}(r)/2\right]$, with the DH potential given by Eq.~(\ref{b42}) of Appendix~\ref{DHsol} in the limit $\kappa_0=\kappa_b$.

First of all, in Fig.~\ref{fig5}, one notes that the DH density profiles for the planar interface and the cylindrical pore coincide, which indicates that curvature effects become negligible for the pore radius $d=10$ nm. Then, it is seen that the DH theory overestimates the MC result for the ion density. The failure of the DH theory for large densities in the range $\rho_{ib}\gtrsim 10^{-2}$ M was shown in Ref.~\cite{jcp2} to result from the unability of the WC theory to account for the local variations of the ionic screening. Finally, the ion densities computed with the present SC scheme is seen in Fig.~\ref{fig5} to exhibit a very good agreement with the MC data. This point shows that the present approach can accurately handle image charge forces beyond the WC regime. Encouraged by this observation, we make use of our SC scheme to reconsider in the next Section~\ref{pht} the ionic CI transition that we had discovered in Refs.~\cite{PRL,jcp1} within a restricted variational scheme.

\begin{figure}
\includegraphics[width=1.1\linewidth]{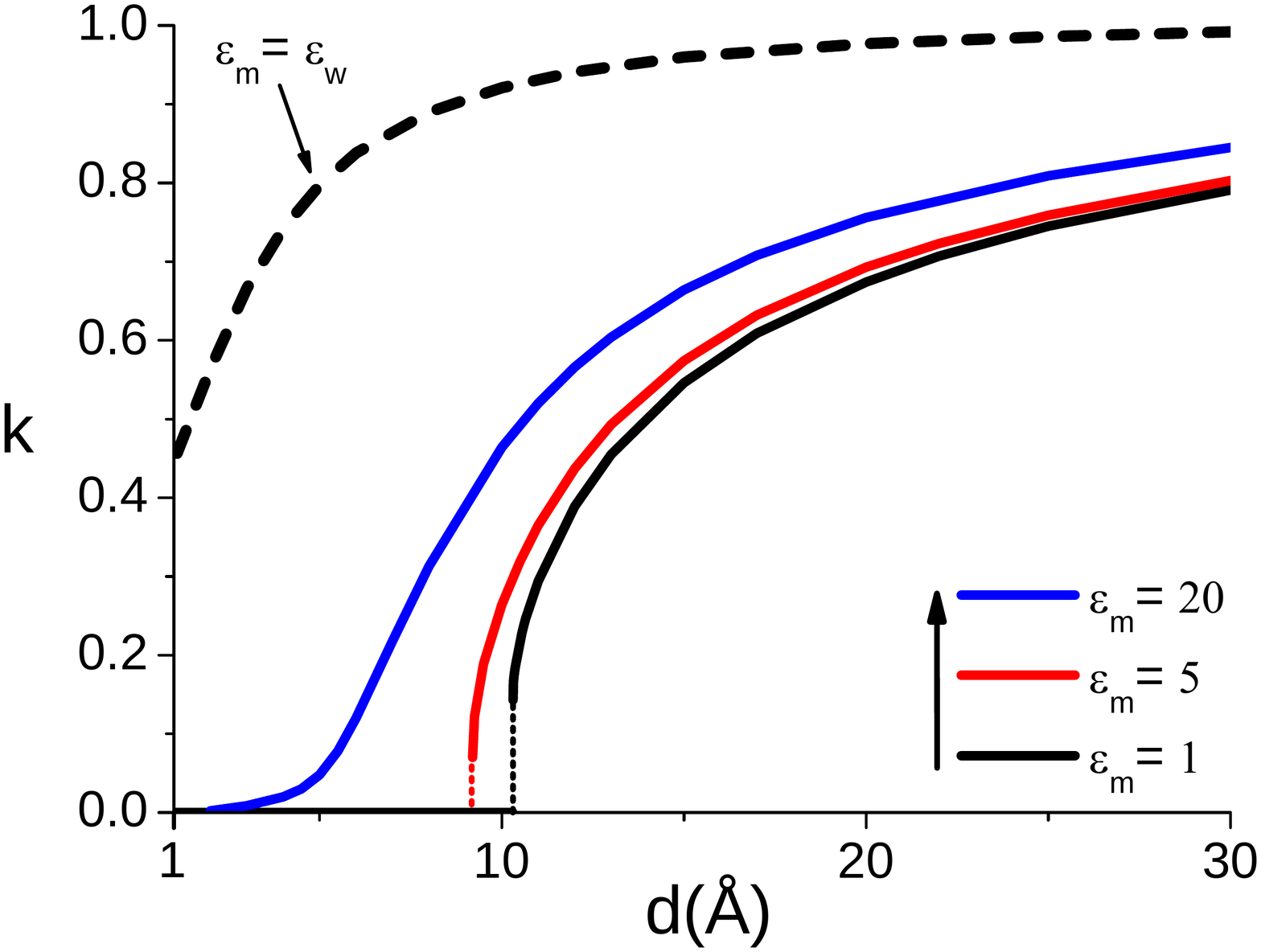}
\caption{(Color online) Ionic partition coefficient against the pore radius for a symmetric electrolyte of two monovalent ion species with reservoir density $\rho_{ib}=0.1$ M and various values of the membrane permittivity $\e_m$. The transition points where the ion density drops to zero are marked by dotted vertical lines.}
\label{fig6}
\end{figure}

\subsubsection{Conductor-insulator transition}
\label{pht}

We illustrate in Fig.~(\ref{fig6}) ionic partition coefficients computed within the present approach from Eqs.~(\ref{defpart}) and~(\ref{denSC}) against the pore radius for different values of the membrane permittivity. In the case of a low permittivity membrane $\e_m=1$ associated with a strong dielectric jump at the pore wall (solid black curve), the partition coefficient is seen to gradually decrease with the pore size, until it sharply drops from $k\simeq\textcolor{black}{0.15}$ to zero at the pore radius $d\simeq$ 10 {\AA}. We note that within the restricted variational scheme based on a uniform pore screening parameter, we had already observed in Refs.~\cite{PRL,jcp1} the same ionic CI transition for cylindrical pores. The appearance of this effect within the general SC scheme of the present study shows that the transition is not a simple artefact of the restricted variational choice in Refs.~\cite{PRL,jcp1}.

Then, with an increase of the membrane permittivity from $\e_m=1$ to $5$, the jump in the ion density is seen to be significantly reduced, and the transition becomes a smooth one for the larger permittivity $\e_m=20$. In Refs.~\cite{PRL,jcp1}, it was shown that the CI transition results from a competition between two opposing effects. On the one hand, the deformation energy of the screening cloud around ions associated with solvation forces is amplified with ionic penetration and thus favors ionic exclusion from the pore. On the other hand, the screening of image interactions that lowers the free energy of the electrolyte favors ionic penetration. The smoothing of the transition stems from the weakening of the competition between these two effects as a consequence of the reduction of image charge forces with increasing $\e_m$. Indeed, the rounding of the transition with a small reduction of the dielectric discontinuity indicates that the effect may be difficult to observe in nanoscale membrane pores where the water permittivity is also expected to be reduced with respect to the reservoir permittivity.

\begin{figure}
(a)\includegraphics[width=1.0\linewidth]{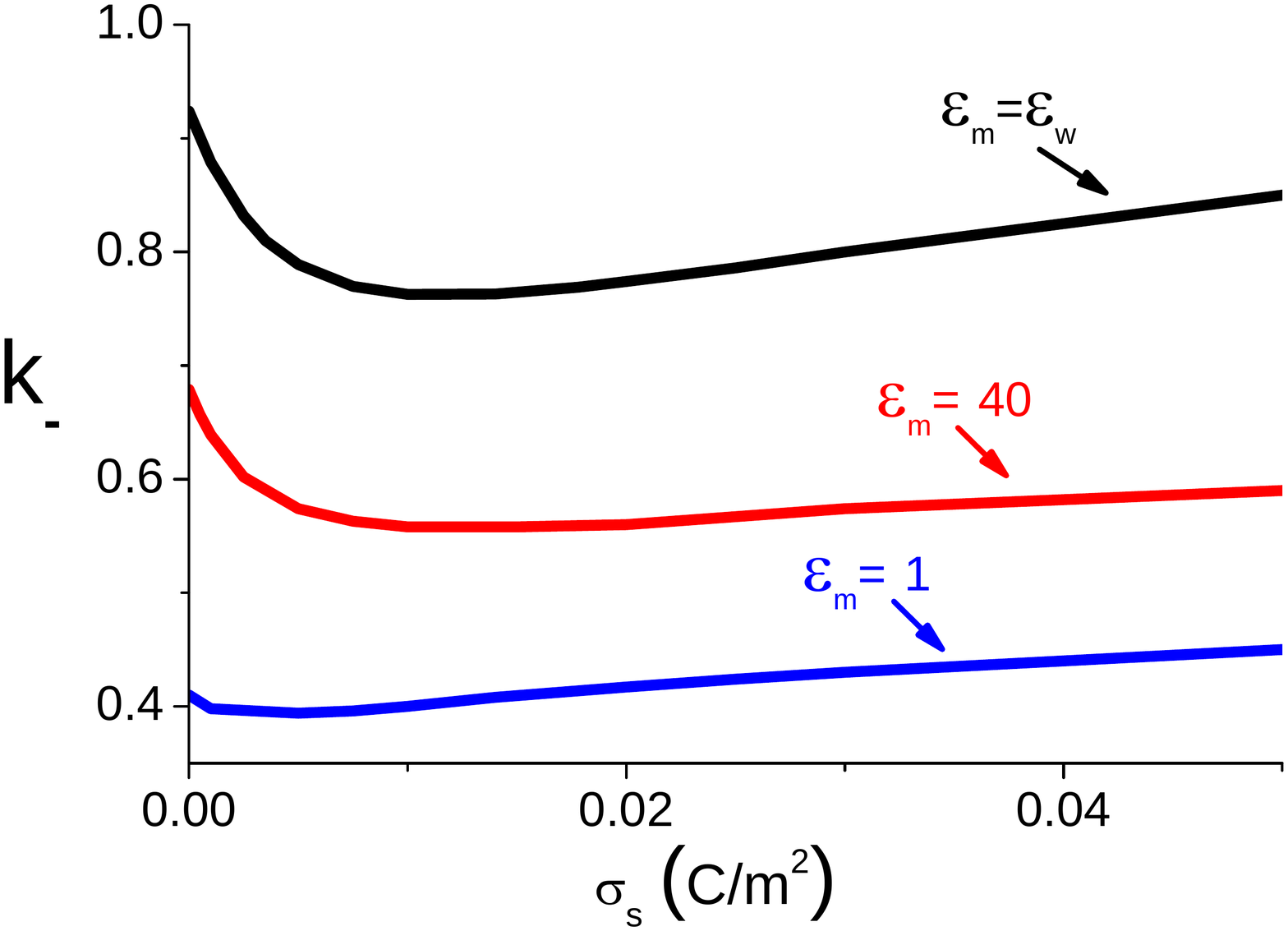}
(b)\includegraphics[width=1.0\linewidth]{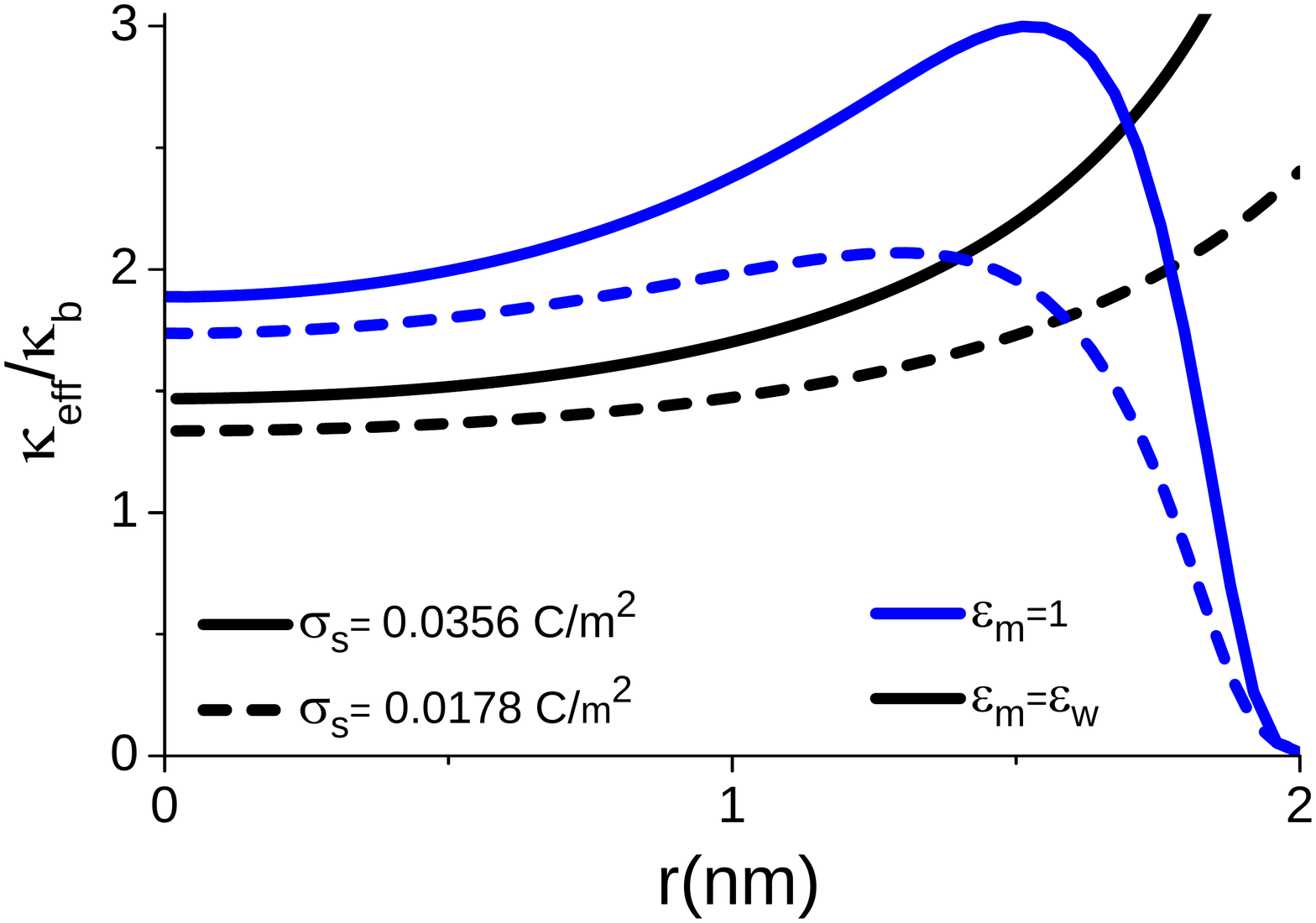}
\caption{(Color online) (a) Coion partition coefficient against the pore charge and (b) the effective screening parameter Eq.~(\ref{kapef}) for an asymmetric electrolyte composed of monovalent coions $q_-=-1$ and divalent counterions $q_+=2$. The bulk anion density is $\rho_{-b}=0.0475$ M and the pore radius $d=2$ nm. In (b), the surface charge values are $\sigma_s=0.0178$ $\mbox{C/m}^2$ (dashed curves) and $\sigma_s=0.0356$ $\mbox{C/m}^2$ (solid curves), and the membrane permittivities $\e_m=1$ (blue curves) and $\e_m=\e_w$ (black curves).}
\label{fig7}
\end{figure}
Finally, in Fig.~\ref{fig6}, the comparison of the lower curves with the result for a dielectrically homogeneous pore $\e_m=\e_w$ shows that regardless of the pore radius, the dielectric exclusion is clearly the dominant rejection mechanism in the nanopore. This is in line with recent nanofiltration experiments where image-charge forces were shown to play the key role in ion rejection from artificial membrane nanopores~\cite{Lang}.

\subsubsection{Enhancement of coion densities with surface charge}
\label{enh}

At the end of Section~\ref{subsec1I} on electrolytes composed of monovalent coions and divalent counterions, it was shown that an increase of the surface charge beyond a characteristic value results in a decrease of the coion rejection rates (see Figs.~\ref{fig3}(a) and (b)). This seemingly counterintuitive effect  displayed in Fig.~\ref{fig7}(a) in terms of the coion partition function (black curve) was shown to result from the strong counterion attraction into the nanopore, transforming the latter into a medium with high screening ability (see the local pore screening parameters displayed by black curves in Fig.~\ref{fig7}(b)).
\begin{figure}
\includegraphics[width=1.0\linewidth]{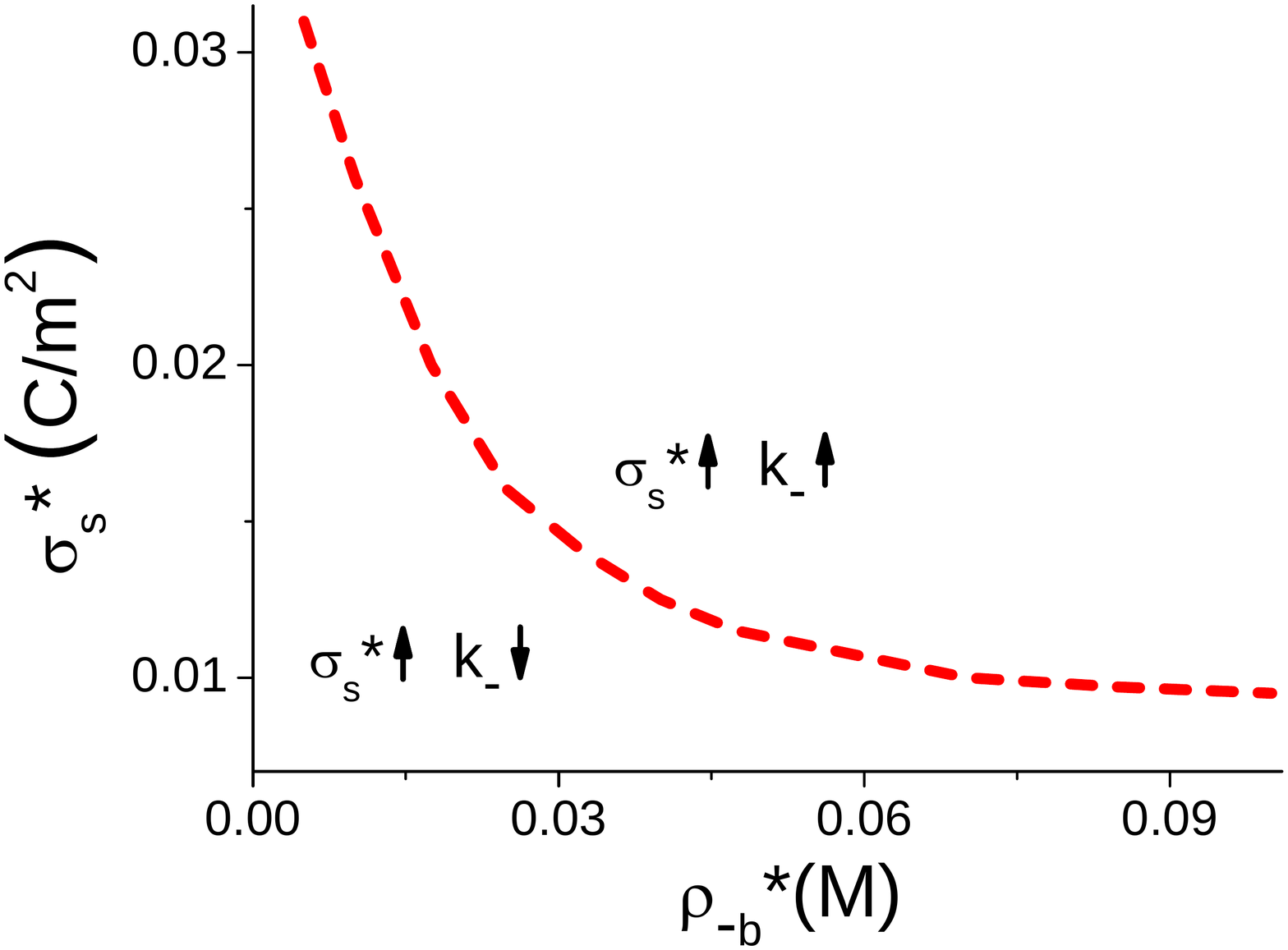}
\caption{(Color online) Diagram characterizing the enhancement of coion densities with the surface charge for an asymmetric electrolyte composed of monovalent coions $q_-=-1$ and divalent counterions $q_+=2$, with the membrane permittivity $\e_m=\e_w$ and the pore radius $d=2$ nm.}
\label{fig8}
\end{figure}

Then, in Fig.~\ref{fig7}(a) where we illustrate the anion partition functions for dielectrically discontinuous pores with $\e_m=1$ and $40$, it is seen that the surface charge induced enhancement of coion densities is also present for low permittivity membrane nanopores characterized by a strong ion rejection. This peculiarity can be explained in terms of the electroneutrality condition in the pore. More precisely, in Fig.~\ref{fig7}(b) where we reported the local screening parameter for $\e_m=1$, the dielectric exclusion of counterions from the pore wall is seen to be compensated by a local counterion excess in the mid-pore area. The latter effect originates from the global electroneutrality condition that fixes the total number of coions and counterions over a crossection of the charged pore. As a result of this mechanism, the enhancement of the coion attraction with the pore charge survives for dielectrically inhomogeneous pores. Thus, the effect should be observable in real membrane nanopores characterized by a low static dielectric permittivity $\e_m=1-5$.

In order to determine the charge density regime where the coion density enhancement is expected, we plot in Fig.~\ref{fig8} the characteristic surface charge $\sigma_s^*$ versus reservoir concentration $\rho^*_{-b}$ that marks the boundary between the area for the surface charge induced anion decrease ($\sigma_s^*\uparrow k_-\downarrow$ below the curve) and increase ($\sigma_s^*\uparrow k_-\uparrow$ above the curve). We note that the transition curve also fixes the applicability limit of the WC DH theory that always predicts an enhanced coion rejection with surface charge, i.e. $\sigma_s^*\uparrow k_-\downarrow$. The diagram in Fig.~\ref{fig8} shows that increasing the ion density from $\rho^*_{-b}=5\times10^{-4}$ M to $0.1$ M, the characteristic pore charge where the effect is expected to come into play is moved from $\sigma_s^*=0.03$ $\mathrm{C/m}^2$ to $0.01$ $\mathrm{C/m}^2$. It should be noted that these values correspond to considerably weak charge densities, and most of the charged nanoscale systems are actually located in the parameter regime above the transition curve of Fig.~\ref{fig8}. For example, the characteristic surface charge of a DNA molecule in a pore or the wall charge of a polyethylene terephthalate membrane nanopore at $\mathrm{pH}=7$ are both on the order $\sigma_s\simeq0.16$ $\mathrm{C/m}^2$. This indicates that the effect scrutinized in this part is relevant to a large variety of biological and industrial nanopore systems, and charge fluctuation effects should be properly taken into account beyond the DH theory for an accurate determination of their functioning.

\section{Conclusion}

In this work, we have characterized electrostatic correlation effects on the ionic selectivity of cylindrical nanopores. To this aim, we have developed an extended one-loop approach that can account for the charge correlations induced by the strong confinement in the cylindrical pore, the surface charge, and the interfacial polarization charges associated with the low dielectric permittivity of the membrane. We have confirmed the quantitative accuracy of the theory and determined its validity regime by making extensive comparisons with MC simulation data.

In Section~\ref{subsec1}, we made use of the present theory in order to analyze MC simulation results of Ref.~\cite{mcasy,mcmix} for the partition of asymmetric and mixed electrolytes in dielectrically homogeneous pores. It was shown that for nanoscale pores with surface charges in the range $\sigma_s\leq 0.0712$ $\mathrm{C/m}^2$, ionic correlation effects can be categorized into two parameter regimes. For electrolytes with monovalent counterions where the size of the surface counterion layer is larger than the screening cloud radius in the bulk, i.e. $\mu\gg\kappa_b^{-1}$, charge screening deficiency close to the pore wall results in an electrostatic barrier decreasing the MF prediction for ion densities in the pore. We emphasize that this effect can be already taken into account in a qualitative way by the weak-coupling DH theory.

In the case of an electrolyte solution with divalent counterions such as $\mathrm{CaCl}_2$ where one gets into the second regime $\mu\ll\kappa_b^{-1}$, the compact interfacial counterion layer responsible for a charge screening excess in the nanopore was shown to attract both negatively and positively charged particles, increasing their pore density above the MF prediction. This second parameter regime is precisely not covered by the DH theory since the latter is unable to account for the non-uniform screening of the electrostatic potential.  Most importantly, we found that beyond a characteristic value of the pore charge where the screening ability of the nanopore overcomes the Donnan rejection mechanism, the coion density rises with the surface charge.

In Section~\ref{subsec2}, we incorporated into this picture image charge interactions associated with the dielectric discontinuity between the pore and the membrane. First of all, it was shown that the general SC approach introduced in this work yields the same ionic CI transition as the one already observed by the restricted variational approach of Refs.~\cite{PRL,jcp1}. This suggests that the transition is not an artefact of the restricted trial potential used in the previous variational theory. However, we also showed that the transition is smoothed if the membrane permittivity exceeds $\e_m\simeq 5$, which indicates that the effect may be difficult to observe in membrane nanopores where the strong confinement is also expected to reduce the pore dielectric permittivity. \textcolor{black}{Thus, in order to consider the case of biological pores with subnanometer radius, future works should consider the electrostatic nature of the solvent beyond the dielectric continuum approximation.} Moreover, the enhancement of coion densities with surface charge in the presence of divalent counterions was shown to survive for dielectrically inhomogeneous pores. Thus, this effect should be observable in biological and artificial nanopores characterized by a low membrane permittivity. Finally, for electrolytes with submolar bulk concentrations, the effect was shown to come into play at considerably weak surface charge densities on the order $\sigma_s\sim 0.01$ $\mathrm{C/m}^2$. This means that the surface charge induced anion density enhancement should be present in many biological and artificial pore systems where the characteristic surface charges are usually an order of magnitude above this limit. This observation confirms the significance of the present SC approach in the study of nanopores with multivalent counterions.

The approach developed herein presents important advantages over the existing theoretical tools of confined electrolytes. First, for dielectrically uniform pores, it allows to determine the partition of the ions in the cylindrical nanopore at one-loop level, which has been so far limited to single planar interfaces~\cite{netzcoun,1loop,jcp2}. Second, the formalism can also accurately consider beyond the WC regime the induced polarization charges resulting from the dielectric jump at the pore wall, a complication that cannot be taken into account by MC simulation technics based on the image charge convention in cylindrical pores. We should also mention that the present theory is less time consuming than MC simulations, since in the most complicated case of a charged nanopore with a strong dielectric discontinuity and divalent counterions (see Fig.~\ref{fig7}), the computation of the ion densities for a given parameter set requires less than 20 minutes on a single 2.7 GHz processor.

We would like to discuss as well the limitations of the present theory. As it was shown in Fig.~\ref{fig3}, in the presence of divalent counterions, the quantitative agreement with MC results worsens for pore charges above $\sigma_s\approx 0.02$ $\mathrm{C/m}^2$. Since the theoretical results overestimate the coion attraction into the nanopore, we think that the absence of HC interactions in the theory may be partly responsible for this failure. The incorporation of HC interactions is also necessary in order to consider concentrated solutions above the characteristic density $\rho_{ib}=0.2$ M where these interactions were shown to come into play even at simple planar interfaces~\cite{jcp2}. \textcolor{black} {For lower ion densities where HC collisions play a minor role, one would expect the finite ion size to modify exclusively the ionic densities in the close vicinity of the pore wall associated with an ion-free Stern layer~\cite{lev2,lev3}. The latter can be easily incorporated into the electrostatic Green's function computed in Appendix~\ref{DHsol}, but we expect this improvement to weakly modify the pore average ion densities for channels with nanoscale radius. Moreover, one should note that unlike the biological channels having a finite length, the nanopore model considered in this work is an infinitely long cylinder. The finiteness of the cylinder length is expected to reduce the strength of correlation effects and particularly the dielectric exclusion effect induced by image charge interactions. In a future work, one could extend the present theory by accounting for the finite nanopore length in the same way as it was done in Ref.~\cite{cyl1} at the DH level}. To conclude, we emphasize that in view of the importance of the cylindrical confinement geometry in nanosystems, the present formalism can find important applications from nanofluidics to artificial nanofiltration technology where a quantitatively accurate consideration of charge correlation and surface polarization effects is still missing.
\\
\acknowledgements  \textcolor{black}{We thank Ralf Blossey for a critical reading of our work and his valuable comments. SB gratefully acknowledges support from} The Academy of Finland through its COMP CoE (no. 251748) and NanoFluid grants and from a postdoctoral grant through the french ANR blanc project `Fluctuations in Structured Coulomb Fluids'.
\smallskip
\appendix
\section{Derivation of the reference potential in cylindrical coordinates}
\label{DHsol}
We derive in this Appendix the reference Green's function that will be needed to solve the closure equations~(\ref{eqt1})-(\ref{eqt3}) derived in Appendices~\ref{1lasym} and~\ref{ex1l}. The confinement geometry corresponds to a cylinder of charge $\sigma_s$ and radius $d$ (see Fig.~\ref{fig0}). We choose the reference Green's function in the form of a DH potential with a general but uniform screening parameter $\kappa_0(r)=\kappa_0\theta(d-r)$, and satisfying the differential equation
\be\label{DH2}
\left[\nabla\e(\br)\nabla-\e(\br)\kappa_0^2(r)\right]v_0(\br,\br')=-\frac{e^2}{k_BT}\delta(\br-\br'),
\ee
where we introduced the piecewise dielectric permittivity profile $\e(\br)=\e_w\theta(d-r)+\e_m\theta(r-d)$.  Exploiting the cylindrical symmetry of the system and inserting the Fourier expansion of the potential
\be\label{expcyl}
v_0(\br,\br')=\sum_{m=-\infty}^\infty\int_{-\infty}^\infty\frac{\mathrm{d}k}{4\pi^2}\tv_0(r,r';k,m)e^{ik(z-z')}e^{im(\theta-\theta')},
\ee
into the relation~(\ref{DH2}), the equation for the Fourier transformed Green's function follows as
\bea\label{DH3}
&&\left\{\frac{1}{r}\partial_rr\e(r)\partial_r-\e(r)\left[\kappa_0^2(r)+k^2+\frac{m^2}{r^2}\right]\right\}\tv_0(r,r';k,m)\nonumber\\
&&=-\frac{e^2}{k_BT}\frac{1}{r}\delta(r-r').
\eea
For the solution of the closure equations~(\ref{eqt1})-(\ref{eqt3}), we exclusively need the solution of Eq.~(\ref{DH3}) for the charge sources inside the cylinder, i.e. for $0<r'<d$. In this case, the homogeneous solution of Eq.~(\ref{DH3}) is indeed given by
\bea\label{hom}
&&\tv_0(r,r';k,m)=C_1I_m(\rho_0 r)\theta(r'-r)\theta(d-r)\\
&&+\left[C_2I_m(\rho_0 r)+C_3K_m(\rho_0 r)\right]\theta(r-r')\theta(d-r)\nonumber\\
&&+C_4K_m(kr)\theta(r-d),\nonumber
\eea
where $I_n(x)$ and $K_n(x)$ stand respectively for the modified Bessel functions of the first and second kind, and we also introduced the auxiliary function $\rho_0=\sqrt{\kappa_0^2+k^2}$.

The integration constants $C_i$ with $1\leq i\leq 4$ in Eq.~(\ref{hom}) are obtained by imposing the boundary conditions associated with the continuity of the potential and the displacement field. The first set of boundary conditions follows by imposing the continuity of the Green's function at the position of the source ion $r=r'$ and at the cylinder wall $r=d$,
\bea\label{b1}
\tv_0(r,r';k,m)\left|_{r\to r'^-}\right.&=&\tv_0(r,r';k,m)\left|_{r\to r'^+}\right. ;\\
\label{b2}
\tv_0(r,r';k,m)\left|_{r\to d^-}\right.&=&\tv_0(r,r';k,m)\left|_{r\to d^+}\right..
\eea
The remaining two boundary conditions are in turn obtained by integrating Eq.~(\ref{DH3}) around $r=r'$ and $r=d$, which yields
\bea\label{b3}
&&\partial_r\tv_0(r,r';k,m)\left|_{r\to r'^-}\right.-\partial_r\tv_0(r,r';k,m)\left|_{r\to r'^+}\right.\nonumber \\
&&=\frac{4\pi\ell_B}{r'};\\
\label{b4}
&&\e(r)\partial_r\tv_0(r,r';k,m)\left|_{r\to d^-}\right.=\e(r)\partial_r\tv_0(r,r';k,m)\left|_{r\to d^+}\right.\nonumber\\
\eea
Imposing the boundary conditions in Eqs.~(\ref{b1})-(\ref{b4}) to the homogeneous solution~(\ref{hom}), the latter finally follows for $r<r'$ in the form
\bea\label{DH5}
\tv_0(r,r';k,m)&=&4\pi\ell_B\left[I_m(\rho_0 r)K_m(\rho_0 r')\right.\\
&&\left.\hspace{.9cm}+F_m(k)I_m(\rho_0 r)I_m(\rho_0 r')\right].\nonumber
\eea
where we defined the function
\bea
F_m(k)=\frac{K_m(kd)K'_m(\rho_0 d)-\eta\gamma K'_m(k d)K_m(\rho_0 d)}{\eta\gamma I_m(\rho_0 d)K'_m(k d)-K_m(k d)I'_m(\rho_0 d)}
\eea
with the parameter $\eta=\e_m/\e_w$ and the function $\gamma=k/\rho_0$. Moreover, we note that the solution of the potential for $r>r'$ follows by interchanging in Eq.~(\ref{DH5}) the variables $r$ and $r'$. Finally, the self energy associated with the Green's function~(\ref{DH5}) follows from Eq.~(\ref{self}) as
\be\label{b42}
\delta v_0(r)=\ell_B(\kappa_b-\kappa_0)+\frac{2\ell_B}{\pi}\sum_{m=-\infty}^\infty\int_0^\infty\mathrm{d}k\;F_m(k)I^2_m(\rho_0 r).
\ee

For the solution of the SC equations in dielectrically discontinuous pores, we note that the self-energy in Eq.~(\ref{b42}) can be separated into a solvation and an image-charge part as $\delta v_0(r)=\delta v_0^{(s)}(r)+\delta v_0^{(im)}(r)$, where
\bea\label{b5}
\delta v_0^{(s)}(r)&=&\ell_B(\kappa_b-\kappa_0)\\
\label{b6}
&&+\frac{2\ell_B}{\pi}\sum_{m=-\infty}^\infty\int_0^\infty\mathrm{d}k\;F^{(s)}_m(k)I^2_m(\rho_0 r);\nonumber\\
\delta v_0^{(im)}(r)&=&\frac{2\ell_B}{\pi}\sum_{m=-\infty}^\infty\int_0^\infty\mathrm{d}k\;F^{(im)}_m(k)I^2_m(\rho_0 r).\nonumber\\
\eea
with the auxiliary functions
\bea\label{b7}
F^{(s)}_m(k)&=&\frac{K_m(kd)K'_m(\rho_0 d)-\gamma K'_m(k d)K_m(\rho_0 d)}{\gamma I_m(\rho_0 d)K'_m(k d)-K_m(k d)I'_m(\rho_0 d)};\nonumber\\
&&\\
\label{b8}
F^{(im)}_m(k)&=&-(1-\eta)\frac{\gamma}{d}K_m(kd)K'_m(kd)\\
&&\times\left[\eta\gamma I_m(\rho_0 d)K'_m(k d)-K_m(k d)I'_m(\rho_0 d)\right]^{-1}\nonumber\\
&&\times\left[\gamma I_m(\rho_0 d)K'_m(k d)-K_m(k d)I'_m(\rho_0 d)\right]^{-1},\nonumber
\eea
One sees that in the limit $\e_m=\e_w$, the function~(\ref{b8}) and the image-charge contribution to the self-energy in Eq.~(\ref{b6}) vanish.

\section{One-loop solution of SC equations}
\label{1lasym}

We explain in this Appendix the one-loop solution of the SC equations~(\ref{eq1}) and~(\ref{eq2}) for ions confined in a dielectrically homogeneous interface i.e. $\e_m=\e_w$, with the surface charge distribution $\sigma(\br)=-\sigma_s\delta(r-d)$. The one-loop approximation consists of an expansion of the SC equations around the MF theory in terms of the Green's function $v(\br,\br')$ and the self-energy $\delta v(\br)$. To this end, we split the external potential into a MF and a fluctuating part as $\phi(\br)=\phi_0(\br)+\phi_1(\br)$, where the MF potential $\phi_0(\br)$ is the solution of the PB equation
\be
\label{eq5}
\nabla\e(\br)\nabla\phi_0(\br)+\frac{e^2}{k_BT}\sum_{i=1}^pn_i(\br)q_i=-\frac{e^2}{k_BT}\sigma(\br),
\ee
where we introduced the MF level ionic number density as
\be\label{numden1}
n_i(\br)=\rho_{ib}e^{-V_w(\br)-q_i\phi_0(\br)}.
\ee
Linearizing the SC equations~(\ref{eq1}) and~(\ref{eq2}) in terms of the Green's function and the excess potential $\phi_1(\br)$, and making use of the MF equation~(\ref{eq5}), one obtains the following equations
\bea
\label{eq6}
&&\nabla\e(\br)\nabla\phi_1(\br)-\frac{e^2}{k_BT}\sum_{i=1}^pn_i(\br)q_i^2\phi_1(\br)=-\frac{e^2}{k_BT}\delta\sigma(\br);\nonumber\\
&&\\
\label{eq7}
&&\nabla\e(\br)\nabla v(\br,\br')-\frac{e^2}{k_BT}\sum_{i=1}^pn_i(\br)q_i^2v(\br,\br')\nonumber\\
&&=-\frac{e^2}{k_BT}\delta(\br-\br'),
\eea
where we introduced in Eq.~(\ref{eq6}) the excess charge density
\be
\delta\sigma(\br)=-\frac{1}{2}\sum_{i=1}^pn_i(\br)q_i^3\delta v(\br).
\ee
With the use of the definition of the Green's function
\be\label{grdef}
\int\mathrm{d}\br_1v^{-1}(\br,\br_1)v(\br_1,\br')=\delta(\br-\br'),
\ee
and the inverse of the one-loop Green's function
\bea\label{1l2}
v^{-1}(\br,\br')&=&\left[-\frac{k_BT}{e^2}\nabla\e(\br)\nabla+\sum_{i=1}^pn_i(\br)q_i^2\right]\delta(\br-\br'),\nonumber\\
\eea
Eq.~(\ref{eq6}) can be inverted as
\be\label{p1l1}
\phi_1(\br)=\int\mathrm{d}\br'v(\br,\br')\delta\sigma(\br').
\ee

In order to evaluate the potential~(\ref{p1l1}), one has to compute the one-loop Green's function by solving Eq.~(\ref{eq7}). Since this equation has no analytical solution in cylindrical coordinates, one has find the solution numerically. To speed up the numerical solution scheme, we opt to solve this equation by choosing as the reference potential the Donnan Green's function, with the corresponding kernel that we introduce in the form
\be\label{inD}
v_D^{-1}(\br,\br')=\left[-\frac{k_BT}{e^2}\nabla\e(\br)\nabla+\frac{\kappa_D^2(\br)}{4\pi\ell_B}\right]\delta(\br-\br'),
\ee
and the effective screening parameter
\be\label{scd}
\kappa_D^2(r)=4\pi\ell_B\sum_{i=1}^p\rho_{ib}q_i^2e^{-q_i\phi_D}\theta(d-r).
\ee
Indeed, the approximation in introducing Eq.~(\ref{inD}) consisted in replacing the MF potential $\phi_0(\br)$ in Eq.~(\ref{numden1}) and~(\ref{1l2}) by the constant Donnan potential $\phi_D$, which is solution of the equation
\be\label{don1}
\sum_{i=1}^p\rho_{ib}q_ie^{-q_i\phi_D}=\frac{2\sigma_s}{d}.
\ee
We note that Eq.~(\ref{don1}) follows by neglecting in Eq.~(\ref{eq5}) the spatial variations of the potential, and integrating the rest of the terms over the cross-section of the channel.

Then, using Eqs.~(\ref{grdef}), (\ref{1l2}), and~(\ref{inD}), the relation~(\ref{eq7}) can be formally inverted as
\be\label{gr1l1}
v(\br,\br')=v_D(\br,\br')+\int\mathrm{d}\br_1v_D(\br,\br_1)\delta n(\br_1)v(\br_1,\br'),
\ee
where we introduced the excess number density
\be\label{deln}
\delta n(\br)=\sum_{i=1}^p\rho_{ib}q_i^2e^{-V_w(\br)}\left[e^{-q_i\phi_D}-e^{-q_i\phi_0(\br)}\right].
\ee
Comparing Eqs.~(\ref{DH2}) and~(\ref{inD}), one notices that the Green's function $v_D(\br,\br')$ is obtained from the potential~(\ref{DH5}) computed in Appendix~\ref{DHsol} by setting $\kappa_0=\kappa_D$.

Moreover, accounting for the cylindrical symmetry and substituting into Eqs.~(\ref{p1l1}) and~(\ref{gr1l1}) the Fourier expansion of the Green's function in the form of Eq.~(\ref{expcyl}), the one loop-potential and the Green's function in Fourier basis takes the form
\bea\label{phi12}
\phi_1(r)&=&\int_0^d\mathrm{d}r_1r_1\tv(r,r_1;0,0)\delta\sigma(r_1);\\
\label{v12}
\tv(r,r';k,m)&=&\tv_D(r,r';k,m)\\
&&+\int_0^d\mathrm{d}r_1r_1\tv_D(r,r_1;k,m)\delta n(r_1)\nonumber\\
&&\hspace{1.35cm}\times\tv(r_1,r';k,m).\nonumber
\eea
We finally note that the one-loop level ion densities given in Eq.~(\ref{den1l}) follow by expanding Eq.~(\ref{iden}) to linear order in $\phi_1(r)$ and $\delta v(r)$.

To evaluate the one-loop potential correction in Eq.~(\ref{phi12}), one has to solve Eq.~(\ref{v12}) by iteration. The first step consists in numerically solving the PB Eq.~(\ref{eq5}) with the boundary conditions
\bea\label{boun}
\phi'_0(0)&=&0;\\
\phi'_0(d^-)&=&\frac{2}{\mu},
\eea
where the parameter $\mu=1/(2\pi q\ell_B\sigma_s)$ is the Gouy-Chapman length. Then, at the first iterative step, the MF potential profile is injected into Eqs.~(\ref{deln}) and~(\ref{v12}), together with the Donnan potential obtained from the solution of Eq.~(\ref{don1}), and the Donnan Green's function from Eq.~(\ref{DH5}) with $\kappa_0=\kappa_D$ used as the input function instead of the function $\tv(r_1,r';\bk,m)$ on the r.h.s. of Eq.~(\ref{v12}). At the next iterative step, the obtained solution for  $\tv(r,r';k,m)$ is reinjected into the r.h.s. of Eq.~(\ref{v12}), and this cycle is continued until self-consistency is achieved. In the end, the converged solution for the Fourier transformed potential $\tv(r,r';k,m)$ is used to evaluate the self-energy $\delta v(r)$ that follows from Eq.~(\ref{v12}) in the form
\bea\label{self1}
\delta v(r)&=&\delta v_D(r)\\
&&+\sum_{m=-\infty}^\infty\int_{-\infty}^\infty\frac{\mathrm{d}k}{4\pi^2}\int_0^d\mathrm{d}r_1r_1\tv_D(r,r_1;k,m)\delta n(r_1)\nonumber\\
&&\hspace{4cm}\times\tv(r_1,r;k,m).\nonumber
\eea
We also note that in Eq.~(\ref{self1}), the self-energy $\delta v_D(r)$ follows from Eq.~(\ref{DH5}) by setting $\kappa_0=\kappa_D$.  Then, the Fourier transformed potential $\tv(r,r';k,m)$ is used with the self-energy~(\ref{self1}) in Eq.~(\ref{phi12}) in order to compute the one-loop correction to the external potential $\phi_1(r)$. Finally, the obtained potentials $\phi_0(r)$, $\phi_1(r)$, and $\delta v(r)$ are inserted in Eq.~(\ref{den1l}) in order to evaluate the ion densities.

\section{Extended one-loop theory for dielectrically inhomogeneous pores}
\label{ex1l}

In this Appendix, we generelize the one-loop scheme introduced in Appendix~\ref{1lasym} to dielectrically discontinuous systems. To this end, we first split the external potential and the self-energy into a solvation contribution and a singular part resulting from image-charge interactions as
\bea\label{sp1}
\phi(\br)&=&\phi_0(\br)+\phi_1(\br);\\
\label{sp2}
\delta v(\br)&=&\delta v^{(im)}(\br)+\delta v^{(s)}(\br),
\eea
where we choose the component $\phi_0(\br)$ as the solution of the modified PB (MPB) equation,
\be\label{sp3}
\nabla\e(\br)\nabla\phi_0(\br)+\frac{e^2}{k_BT}\sum_{i=1}^pq_in_i(\br)=-\frac{e^2}{k_BT}\sigma(\br),
\ee
with the number density
\be\label{sp4}
n_i(\br)=\rho_{ib}e^{-V_w(\br)-q_i\phi_0(\br)}e^{-\frac{q_i^2}{2}\delta v^{(im)}(\br)}.
\ee
The image-charge potential $\delta v^{(im)}(\br)$ in Eqs.~(\ref{sp2})-(\ref{sp4}) will be introduced below.

Inserting the potentials in Eqs.~(\ref{sp1})-(\ref{sp2}) into the SC Eqs.~(\ref{eq1})-(\ref{eq2}), linearizing the latter in terms of the potentials $\phi_1(\br)$ and $\delta v^{(s)}(\br)$, and making use of the MPB equation~(\ref{sp3}), one gets the following relations,
\bea\label{sp5}
&&\nabla\e(\br)\nabla\phi_1(\br)-\frac{e^2}{k_BT}\sum_{i=1}^pq_i^2n_i(\br)\phi_1(\br)=-\frac{e^2}{k_BT}\delta\sigma(\br);\nonumber\\
&&\\
\label{sp6}
&&\nabla\e(\br)\nabla v(\br,\br')-\frac{e^2}{k_BT}\sum_{i=1}^pq_i^2n_i(\br)v(\br,\br')\nonumber\\
&&=-\frac{e^2}{k_BT}\delta(\br-\br'),
\eea
where we introduced in Eq.~(\ref{sp5}) the excess charge density
\be\label{anan}
\delta\sigma(\br)=-\frac{1}{2}\sum_{i=1}^pq_i^3n_i(\br)\delta v^{(s)}(\br).
\ee
Identifying from Eq.~(\ref{sp6}) the kernel associated with the Green's function
\bea\label{spl7}
v^{-1}(\br,\br')&=&\left[-\frac{k_BT}{e^2}\nabla\e(\br)\nabla+\sum_{i=1}^pq_i^2n_i(\br)\right]\delta(\br-\br'),\nonumber
\eea
and using the definition of the Green's function in Eq.~(\ref{grdef}), Eq.~(\ref{sp5}) can be inverted to get the correction to the MPB potential as
\be\label{spl8}
\phi_1(\br)=\int\mathrm{d}\br'v(\br,\br')\delta\sigma(\br').
\ee

The evaluation of the potential correction in Eq.~(\ref{spl8}) requires again the knowledge of the Green's function, i.e. the solution of Eq~(\ref{sp6}). We opt to solve this equation in a similar way as in Appendix~\ref{ex1l}, but by choosing now as the reference potential the DH Green's function, whose kernel is obtained by setting in Eq.~(\ref{inD}) the Donnan potential to zero, that is $\psi_D=0$. This gives
\bea\label{inDH}
v_{DH}^{-1}(\br,\br')&=&\left[-\frac{k_BT}{e^2}\nabla\e(\br)\nabla+\sum_{i=1}^p\rho_{ib}q_i^2e^{-V_w(\br)}\right]\delta(\br-\br').\nonumber
\eea
Making use of the relations~(\ref{spl7}), (\ref{inDH}), and the definition of the Green's function~(\ref{grdef}), Eq.~(\ref{sp2}) can be inverted to give
\be\label{gr1l2}
v(\br,\br')=v_{DH}(\br,\br')+\int\mathrm{d}\br_1v_{DH}(\br,\br_1)\delta n(\br_1)v(\br_1,\br'),
\ee
with the excess number density
\be\label{spl9}
\delta n(\br)=\sum_{i=1}^p\rho_{ib}q_i^2e^{-V_w(\br)}\left[1-e^{-q_i\phi_0(\br)-\frac{q_i^2}{2}\delta v^{(im)}(\br)}\right].
\ee
Taking into consideration the cylindrical asymmetry and substituting into Eqs.~(\ref{spl8}) and~(\ref{gr1l2}) the Fourier expansion of the Green's function~(\ref{expcyl}), one finally gets
\bea\label{spl10}
\phi_1(r)&=&\int_0^d\mathrm{d}r_1r_1\tv(r,r_1;0,0)\delta\sigma(r_1);\\
\label{spl11}
\tv(r,r';k,m)&=&\tv_{DH}(r,r';k,m)\\
&&+\int_0^d\mathrm{d}r_1r_1\tv_{DH}(r,r_1;k,m)\delta n(r_1)\nonumber\\
&&\hspace{1.35cm}\times\tv(r_1,r';k,m).\nonumber
\eea
The Fourier-transformed DH potential in Eq.~(\ref{spl11}) follows from Eq.~(\ref{DH5}) by taking the limit $\kappa_0=\kappa_b$.

In order to solve Eqs.~(\ref{sp3}) and~(\ref{spl10})-(\ref{spl11}), we have to define the image charge part of the self-energy in the exponential of the excess density function~(\ref{spl9}). To this aim, we first split the latter into a solvation and and image part as
\be\label{spl12}
\delta n(r)=\delta n^{(s)}(r)+\delta n^{(im)}(r),
\ee
with
\bea\label{spl13}
\delta n^{(s)}(r)&=&\sum_{i=1}^p\rho_{ib}q_i^2e^{-V_w(r)-\frac{q_i^2}{2}\delta v^{(im)}(r)}\left[1-e^{-q_i\phi_0(r)}\right];\nonumber\\
&&\\
\label{spl14}
\delta n^{(im)}(r)&=&\sum_{i=1}^p\rho_{ib}q_i^2e^{-V_w(r)}\left[1-e^{-\frac{q_i^2}{2}\delta v^{(im)}(r)}\right].
\eea
Injecting the decomposition in Eq.~(\ref{spl12}) into Eq.~(\ref{gr1l2}) and taking the limit $\br'=\br$, one can recast the equal point Green's function in the form $\delta v(r)=\delta v^{(s)}(r)+\delta v^{(im)}(r)$, where the solvation and image-charge contributions are given by
\bea\label{spl15}
\delta v^{(\alpha)}(r)&=&\delta v_{DH}^{(\alpha)}(r)\\
&&+\sum_{m=-\infty}^\infty\int_{-\infty}^\infty\frac{\mathrm{d}k}{4\pi^2}\int_0^d\mathrm{d}r_1r_1\tv_{DH}(r,r_1;k,m)\nonumber\\
&&\hspace{3.2cm}\times\delta n^{(\alpha)}(r_1)\tv(r_1,r;k,m),\nonumber
\eea
with the superscript $\alpha=\{im,s\}$. We finally note that solvation and polarization parts of the DH self-energy in Eq.~(\ref{spl15}) are obtained from Eqs.~(\ref{b5}) and~(\ref{b6}) by setting $\kappa_0=\kappa_b$.

Because the image-charge potential in Eq.~(\ref{spl15}) is considered self-consistently, the solution of the closure equations composed of the relations~(\ref{sp3}), (\ref{spl10})-(\ref{spl11}), and the relation Eq.~(\ref{spl15}) for the image potential is more tricky. Describing the iterative solution, we will skip some of the details already explained in Appendix~\ref{1lasym}. The solution scheme consists in numerically solving first the MPB equation~(\ref{sp3}) by using as the image potential $\delta v_{im}(r)$ the dielectric part of the weak-coupling DH potential $\delta v^{(0)}_{im}(r)$ in Eq.~(\ref{b6}) with $\kappa_0=\kappa_b$, i.e. $\delta v_{im}(r)\to\delta v^{(0)}_{im}(r)$. At the next iterative step, the obtained solution for $\phi_0(r)$ and the DH potential $\delta v^{(0)}_{im}(r)$ are used in Eqs.~(\ref{spl9}) and ~(\ref{spl11}) in order to compute the output function $\tv(r,r';k,m)$. The latter is injected in turn into Eq.~(\ref{spl15}) in order to obtain the new values of the image potential $\delta v_{im}(r)$. Then, the MPB equation~(\ref{sp3}) is numerically solved with the updated image potential $\delta v_{im}(r)$, which provides us with the updated external potential $\phi_0(r)$. One has to continue the cycle until the converged Green's function $\tv(r,r';k,m)$ is obtained. Finally, using the latter in Eqs.~(\ref{spl15}),  (\ref{anan}), and~(\ref{spl10}), one gets the external potentials $\phi_0(r)$, $\phi_1(r)$, and the self-energies $\delta v_{s}(r)$ and $\delta v_{im}(r)$ that allows to evaluate the ion density  in Eq.~(\ref{denSC}).


\begin{thebibliography}{99}
\bibitem {yar1} A.E. Yaroshchuk, Adv. Colloid Interf. Sci. \textbf{85}, 193 (2000).
\bibitem {yar2} A. Yaroshchuk, Sep. Purif. Technology 22-23, 143 (2001).
\bibitem {Lang}  C.-H. Tu, Y.-Y. Fang, J. Zhu, B. Van der Bruggen, and X.-L. Wang, Langmuir \textbf{27}, 10274 (2011).
\bibitem {nano1} H. Daiguji, P.D. Yang, A.J. Szeri, and A.J. Majumdar, Nano Lett. \textbf{4}, 2315 (2004).
\bibitem {nano2} S.L. Levy and H.G. Craighead, Chem. Soc. Rev. \textbf{39}, 1133 (2010).
\bibitem {nano3} \textit{Lab-on-a-Chip Technology : Biomolecular Separation and Analysis}, edited by K.E. Herold, and A. Rasooly, Caister Academic Press (2009).
\textcolor{black}{
\bibitem {dun1} Rob D. Coalson,  A. Duncan and N. B. Tal, J. Phys. Chem. \textbf{100}, 2612 (1996).
\bibitem {orland1} A. Abrashkin, D. Andelman, and H. Orland, Phys. Rev. Lett. \textbf{99}, 077801 (2007).
\bibitem {buyukdagli13} S. Buyukdagli and T. Ala-Nissila, Phys. Rev. E \textbf{87}, 063201 (2013).
\bibitem {jcpPOL} S. Buyukdagli  and T. Ala-Nissila, J. Chem. Phys. \textbf{139}, 044907 (2013).
\bibitem {dun2} S. Tsonchev,  R.D. Coalson, and A. Duncan, Phys. Rev. E \textbf{76}, 041804 (2007).}
\bibitem {mcasy} B. Jamnik and V. Vlachy, J. Am. Chem. Soc. \textbf{115}, 660 (1993).
\bibitem {mcmix} B. Jamnik and V. Vlachy, J. Am. Chem. Soc. \textbf{117}, 8010 (1995).
\bibitem {SCrussian} S.M. Avdeev, G.A. Martynov, Colloid J. USSR \textbf{48}, 632 (1986).
\bibitem {netzvar} R.R. Netz and H. Orland, Eur. Phys. J. E \textbf{11}, 301 (2003).
\textcolor{black}{
\bibitem {anan1} H.-K. Tsao, J. Phys. Chem. B \textbf{102}, 10243 (1998).
\bibitem {anan2} K. Bohinc, A. Iglic, T. Slivnik, and V. Kralj-Iglic, Cell. Mol. Biol. Lett. \textbf{7}, 839 (2002).
\bibitem {anan3}  K. Bohinc, J. Gimsa, V. Kralj-Iglic, T. Slivnik, and A. Iglic, Bioelectrochemistry \textbf{67}, 91 (2005).}
\bibitem {PodWKB} R. Podgornik and B. Zeks, J. Chem. Soc. Faraday Trans. 2 \textbf{84}, 611 (1988).
\bibitem {attard} P. Attard, D.J. Mitchell, and B.W. Ninham, J. Chem. Phys. \textbf{89}, 4358 (1988).
\bibitem {netzcoun}  R.R. Netz and H. Orland, Eur. Phys. J. E \textbf{1}, 203 (2000).
\bibitem {1loop} A. W. C. Lau, Phys. Rev. E \textbf{77}, 011502 (2008).
\bibitem {jcp2} S. Buyukdagli, C.V. Achim, and T. Ala-Nissila, J. Chem. Phys. \textbf{137}, 104902 (2012).
\bibitem {st1} B.I. Shklovskii, Phys. Rev. E, \textbf{60}, 5802 (1999).
\bibitem {st2} A. G. Moreira and R. R. Netz,  Europhys. Lett., \textbf{52} (6), 705 (2000) .
\bibitem {st3} O. Punkkinen, A. Naji, R. Podgornik, I. Vattulainen, and P.L. Hansen, Europhys. Lett. \textbf{82}, 48001 (2008).
\bibitem {st4} Y. S. Jho, M. Kanduc, A. Naji, R. Podgornik, M.W. Kim, and P. A. Pincus, Phys. Rev. Lett. \textbf{101}, 188101 (2008).
\bibitem {cyl1} Y. Levin, Europhys. Lett., \textbf{76} , 163 (2006).
\bibitem {cyl2} D.S. Dean and R.R. Horgan, J. Phys. C. \textbf{17}, 3473, (2005).
\bibitem {cyl3} M. Kanduc, A. Naji and R. Podgornik, J. Chem. Phys., \textbf{132}, 224703 (2010).
\bibitem {David} D. S. Dean and R. R. Horgan, Phys. Rev. E \textbf{70}, 011101 (2004).
\bibitem {japan} T. Nakamura, T. Tanaka and Y. Izumitani, J. Phys. Soc. Jpn. \textbf{51}, 2271 (1982).
\bibitem {hatlo} M.M. Hatlo, R.A. Curtis and L. Lue, J. Chem. Phys. \textbf{128}, 164717 (2008).
\bibitem {pre} S. Buyukdagli, M. Manghi, and J. Palmeri, Phys. Rev. E \textbf{81}, 041601 (2010).
\bibitem {PRL} S. Buyukdagli, M. Manghi, and J. Palmeri, Phys. Rev. Lett. \textbf{105}, 158103 (2010).
\bibitem {jcp1} S. Buyukdagli, M. Manghi, and J. Palmeri, J. Chem. Phys. \textbf{134} 074706 (2011).
\bibitem {exp0} K.A. Kraus, A.E. Marcinkowsky, J.S. Johnson, A.J. Shor, Science \textbf{151}, 194 (1966).
\textcolor{black}{\bibitem {lev2} Y. Levin and J.E. Flores-Mena, Europhys. Lett., \textbf{56} , 187 (2001).
\bibitem {lev3} A. Bakhshandeh, A.P. Dos Santos, and Y. Levin, Phys. Rev. Lett. \textbf{107}, 107801 (2011).}


\end{thebibliography}
\end{document}